\documentclass[useAMS,usenatbib]{mn2e}

\usepackage{psfig, epsf, epsfig}

\title[Star formation in groups and clusters]{Galactic star formation enhanced and quenched
by ram pressure in groups and clusters}
\author[K. Bekki]
{Kenji Bekki${}^1$\thanks{E-mail:
bekki@cyllene.uwa.edu.au} \\
${}^1$ICRAR M468
The University of Western Australia
35 Stirling Hwy, Crawley
Western Australia 6009, Australia}

\begin{document}

\date{Accepted, Received 2005 February 20; in original form }

\pagerange{\pageref{firstpage}--\pageref{lastpage}} \pubyear{2005}

\maketitle

\label{firstpage}

\begin{abstract}

We investigate how ram pressure of intragroup and intracluster 
medium can influence the spatial and temporal variations of
star formation (SF)  of disk galaxies 
with halo masses ($M_{\rm h}$) ranging from
$10^{10} {\rm M}_{\odot}$ to $10^{12} {\rm M}_{\odot}$ (i.e., from dwarf irregular
to Milky Way-type) 
in groups and clusters with 
$10^{13} \le M_{\rm h}/{\rm M}_{\odot}  \le 10^{15}$
by using numerical  simulations with a new model for time-varying ram pressure.
The long-term evolution of SF rates and H$\alpha$ morphologies 
corresponding to the distributions of star-forming regions are particularly investigated
for different model parameters. The principal results are as follows.
Whether ram pressure can enhance or reduce SF  depends on 
$M_{\rm h}$ of disk galaxies and 
inclination angles of gas disks with respect to their orbital directions 
for a given orbit and a given environment.
For example,  SF 
can be moderately  enhanced in disk galaxies
with $M_{\rm h}=10^{12} {\rm M}_{\odot}$ at the pericenter passages 
in a cluster with $M_{\rm h}=10^{14} {\rm M}_{\odot}$ whereas
it can be completely shut down (`quenching') for low-mass disks 
with $M_{\rm h}=10^{10} {\rm M}_{\odot}$.
Ram pressure can reduce the H$\alpha$-to-optical-disk-size ratios of disks 
and the revel of the reduction depends on $M_{\rm h}$ and orbits of disk galaxies
for a given environment.
Disk galaxies under strong ram
pressure show characteristic H$\alpha$ morphologies such as ring-like, one-sided, and  
crescent-like distributions. 
\end{abstract}

\begin{keywords}
galaxies: clusters : general --
galaxies:ISM --
galaxies:evolution --
stars:formation  
\end{keywords}

\section{Introduction}

Galaxy evolution in groups and clusters can be significantly influenced
by environmental processes, such as tidal compression of cluster gravitational
fields (e.g., Byrd \& Valtonen 1990),
merging between group member galaxies (e.g., Mamon 2001),
ram pressure stripping of disk gas (e.g., Gunn \& Gott 1972),
high external pressure of intracluster medium (ICM) 
on galactic gas disks (e.g., Evrard 1991),
multiple high-speed galaxy encounters (e.g., Moore et al. 1994),
and stripping of galactic halo gas by ram pressure of ICM (e.g., Bekki 2009).
One of key issues in extragalactic astronomy is to understand the relative
importance of each of these environmental processes in galaxy evolution
for different environments (See Boselli \& Gavazzi 2006 for a recent review on
this). Ram pressure effect on galactic gas disk 
is distinct from other environmental processes
based on gravitation interaction
in the sense that it can influence only the cold gaseous components of disk galaxies
(i.e., it can not directly influence the stellar components).
Ram pressure of ICM, however, can be a key process for better understanding
the star formation histories (SFHs) of disk galaxies in groups and clusters,
as described throughout this paper.

Recent observational studies have shown that ram pressure stripping 
can influence temporal and spatial variations
of star formation (SF) in disk galaxies (e.g., Koopmann \& Kenny 2004; Crowl \& Kenny 2008;
Cortese et al. 2012; Fossati et al. 2013). 
For example, Koopmann \& Kenny (2004) found that spiral galaxies with truncated
H$\alpha$ disks show undisturbed stellar disks in the Virgo cluster and accordingly
suggested that  hydrodynamical interaction between ICM and 
interstellar medium (ISM) is responsible
for the truncated H$\alpha$ disks.
Also, these observations found that the H$\alpha$-to-optical-disk-size ratio
is smaller  for spiral galaxies with higher degrees of H~{\sc i}-deficiency.
(e.g., a negative correlation between $r_{\rm e}({\rm H}_{\alpha})/r_{\rm e}$
and Def$_{\rm HI}$; Fossati et al. 2013). 
Given that the time evolution of global colors
and chemical abundances of disk galaxy depends strongly on SFHs,
these observations  suggest that understanding the roles of ram pressure
in SFHs of galaxies is important also for better understanding galaxy evolution
in general.

Although many numerical simulations investigated in what physical conditions
ISM of galaxies with different morphological types
can be stripped by ram pressure
(e.g., Abadi et al. 1999; Mori \& Burkert 2000;
Quilis et al. 2000; Schultz \& Struck 2001; Vollmer et al. 2001;
Roediger \& Hensler 2005; Roediger \& Br\"uggen 2007;
J\'achym et al. 2007; Kronberger et al. 2008; Tonnesen \& Bryan  2012),
a surprisingly small number of simulations
have so far investigated how the SFHs of disk galaxies
can be influenced by ram pressure.
Bekki \& Couch (2003) demonstrated that external high pressure of ICM on gas clouds in disk
galaxies can induce
the gravitational collapse of the  clouds and thus can trigger a very efficient star formation
(starburst).
Kronberger et al. (2008) showed that ram pressure of ICM can significantly enhance
(by up to a factor of 3) the mean star formation rates in the central regions of 
disk galaxies, if the ICM has a density of $10^{-28}$ g cm$^{-3}$ and 
a temperature of $3.6 \times 10^7$ K.
Mastropietro et al. (2009) investigated how  ram pressure of the Galactic halo gas can influence
the distribution of star-forming regions in the Large Magellanic Cloud (LMC).
Tonnesen \& Bryan (2012) demonstrated that star formation in galactic bulges
can be slightly enhanced in disk galaxies under moderately strong ram pressure of ICM.

Although these previous simulations clearly demonstrated that external high pressure 
of ICM can possibly
enhance star formation in disk galaxies under some physical conditions, 
they did not investigate the long-term (more than a few Gyr) SFHs of galaxies under
ram pressure  in 
a systematic parameter study. 
Furthermore, these previous simulations did not properly model
the time evolution of the strength of ram pressure  for disk galaxies which
enter into clusters from fields.
Since the strength of ram pressure depends both on (i) the relative velocities
between ICM and galaxies and on (ii) the densities of ICM surrounding galaxies,
orbital evolution of galaxies within clusters needs to be included
in simulating the long-term effects of ram pressure on galaxies.
Accordingly, it is unclear how ram pressure can influence 
SFHs of galaxies  when they enter into a cluster environment and passes
through its central region. A more comprehensive systematic parameter study on the influences
of ram pressure on SFHs of galaxies is necessary.

The purpose of this paper is to investigate how ram pressure of ICM can influence
the long-term SFHs 
of disk galaxies in a comprehensive manner
by using numerical  simulations with a proper model for time-varying ram pressure
force.  We particularly investigate
the following questions: (i) whether ram pressure can enhance or reduce star formation
in disk galaxies, (ii) how the effects of ram pressure on galactic SFHs
depend on galaxy properties and environments, 
(iii) how the H$\alpha$ distributions of disk galaxies under strong ram pressure look like,
(iv) how the ratios of H$\alpha$ disk sizes to optical disk ones depend on 
galaxy/group/cluster masses,
and (v) how ram pressure is important for better understanding the observed properties
of galaxies in groups and clusters.

Given ongoing extensive observational studies of 2D H$\alpha$ distribution
of star-forming regions in galaxies by multi-object spectrograph on large ground-based
telescopes (e.g., Croom et al. 2012; Brough et al. 2013), 
it is very timely for the present study to
predict 2D H$\alpha$ distributions of star-forming regions in disk galaxies under strong ram
pressure. These predictions will help observers to interpret the origin of the H$\alpha$
morphologies of disk galaxies in different environments 
(e.g., Moss \& Whittle 2000;  Koopmann \& Kenny 2004).
Galactic SFHs in groups and clusters have been discussed in the context
of gradual decline due to {\it halo gas}
 stripping (`strangulation'; Larson et al. 2002;
Balogh et al. 2000; Bekki et al. 2001, 2002; Shioya et al. 2004).
Ram pressure stripping of {\it disk gas}
can cause sudden and rapid removal of gas and thus can have more dramatic effects
on SFHs of galaxies in comparison with the strangulation mechanism.

The plan of the paper is as follows: In the next section,
we describe our new   model for time-dependent ram pressure force on disk galaxies
in clusters.
In \S 3, we
present the numerical results
on the long-term SFHs and spatial distributions of SF regions 
in disk galaxies under strong ram pressure.
In this section, we also discuss the dependences of the results on the adopted
model parameters.
In \S 4, we discuss the possibly important implications
of the  simulations results.
We summarize our  conclusions in \S 5.

\begin{table*}
\centering
\begin{minipage}{80mm}
\caption{Description of the basic parameter values
for the three different disk galaxy  models.}
\begin{tabular}{lccc}
Model name/Physical properties
& MW-type  & M33-type  & Dwarf-type \\
DM mass ($\times 10^{12} {\rm M}_{\odot}$) & 1.0 & 0.1 & 0.01 \\
Virial radius  (kpc) & 245.0  & 113.8 & 52.9 \\
{$c$  \footnote{$c$ is the $c$-parameter in the NFW dark matter
profiles.}}
&  10 & 12 & 16 \\
Stellar disk  mass ($\times 10^{10} {\rm M}_{\odot}$) & 5.4 & 0.36 & 0.036 \\
Gas disk  mass ($\times 10^{10} {\rm M}_{\odot}$) & 1.2 & 0.3 & 0.03 \\
Bulge  mass ($\times 10^{10} {\rm M}_{\odot}$) & 1.0 & -- & -- \\
Stellar disk size (kpc) & 17.5 & 8.1 & 3.8 \\
Gas disk size (kpc)  & 35.0 & 16.2 & 7.6 \\
Bulge  size (kpc) & 3.5 & -- & -- \\
\end{tabular}
\end{minipage}
\end{table*}

\section{The model}

A disk galaxy  is assumed to move
in a cluster with a total halo mass $M_{\rm h}$ and therefore be influenced
by ram pressure of the ICM. The ram pressure force on the disk galaxy
is calculated according to the position and velocity of the galaxy
with respect to the cluster center. The tidal field of the cluster
and tidal interaction with other cluster member galaxies are not included
in the present study
so that ram pressure effects on galactic SFHs can be more clearly investigated.
Other key environmental effects on galaxies (e.g., slow and fast tidal encounters)
will be investigated in our forthcoming papers.
We focus exclusively on how SFHs of disk galaxies under strong ram pressure of ICM
depend on galaxy properties and environments.

In order to simulate the time evolution of
SFRs and gas contents in disk galaxies  under ram pressure of ICM,
we use our original chemodynamical simulation code that can be run
on GPU machines (Bekki 2013).
The code is a revised version of our original GRAPE-SPH code (Bekki 2009)
which combines the method of smoothed particle
hydrodynamics (SPH) with GRAPE for calculations of three-dimensional
self-gravitating fluids in astrophysics. 
This code enable us to investigate the formation and evolution of dust and molecular
hydrogen (${\rm H}_2$) in disk galaxies (Bekki 2013), though it is numerically
costly to calculate ${\rm H}_2$ formation and evolution
(i.e., it requires much longer CPU time in comparison with our simulations with no
${\rm H}_2$ calculation).
Since our main interest is the time evolution of SFRs in galaxies (not the evolution
of dust and ${\rm H}_2$ contents),  we do not incorporate ${\rm H}_2$ calculations
into the present simulation.

\subsection{A disk galaxy}

A disk  galaxy  is composed of  dark matter halo,
stellar disk,  stellar bulge, and  gaseous disk, as is assumed in
other numerical simulations of ram pressure stripping (e.g., Abadi et al 1999;
J\'achym et al. 2007). 
Our previous study (Bekki 2009) investigated already the ram pressure stripping
of halo gas in disk galaxies within groups and clusters of galaxies.
We therefore do not investigate the ram pressure stripping
of halo gas by modeling halo gas in the present simulation.
The total masses of dark matter halo, stellar disk, gas disk, and
bulge are denoted as $M_{\rm h}$, $M_{\rm s}$, $M_{\rm g}$,
and $M_{\rm b}$, respectively. The total disk mass (gas + stars)
and gas mass fraction
are denoted as $M_{\rm d}$ and $f_{\rm g}$, respectively,  for convenience.
The mass ratio of the dark matter halo ($M_{\rm h}$) to the  disk 
($M_{\rm s}+M_{\rm g}$) 
in a disk galaxy  is fixed at 16.7.

We mainly investigate three different disk galaxy models with different
$M_{\rm h}$ in order to clarify how a galaxy environment can influence different
types of galaxies in a different manner. The three includes the `MW-type'
(labeled as `MW' for simplicity)
with $M_{\rm h}=10^{12} {\rm M}_{\odot}$,
`M33-type' (`M33')
with $M_{\rm h}=10^{11} {\rm M}_{\odot}$,
and `Dwarf-type' (`DW'),
with $M_{\rm h}=10^{10} {\rm M}_{\odot}$. 
These three disk models have physical properties which are similar to
yet not exactly the same as those
observed for the MW, M33, and typical dwarfs. These name are used just for 
distinguishing three disk galaxies with different $M_{\rm h}$. The MW-type
galaxy has a bulge and a smaller gas mass fraction whereas the other two
have no bulge and a larger gas mass fraction.
The basic parameter values for these three disk models are summarized in
Table 1.

\begin{table*}
\centering
\begin{minipage}{180mm}
\caption{Description of the basic parameter values
for the three different disk galaxy  models.}
\begin{tabular}{lccc}
Model/Physical properties
& {Group \footnote{These definition of `Group', `Low-mass cluster', and `High-mass cluster'
by mass ($M_{\rm h}$) would be arbitrary.}}
&  Low-mass cluster (`Virgo')
& High-mass cluster (`Coma')  \\
Total mass ($\times 10^{14} {\rm M}_{\odot}$) & 0.1 & 1.0 & 10.0 \\
Virial radius  (Mpc) & 0.56 & 1.20  & 2.59 \\
{$c$  \footnote{$c$ is the $c$-parameter in the NFW dark matter
profiles.}}
& 6.0 &  4.9 & 3.6 \\
ICM  mass ($\times 10^{14} {\rm M}_{\odot}$) & 0.015  & 0.15 & 1.5 \\
ICM  temperature  ($\times 10^{7}$ K) & 0.56 & 2.6 & 12.0 \\
\end{tabular}
\end{minipage}
\end{table*}

We adopt the density distribution of the NFW
halo (Navarro, Frenk \& White 1996) suggested from CDM simulations:
\begin{equation}
{\rho}(r)=\frac{\rho_{0}}{(r/r_{\rm s})(1+r/r_{\rm s})^2},
\end{equation}
where  $r$, $\rho_{0}$, and $r_{\rm s}$ are
the spherical radius,  the characteristic  density of a dark halo,  and the
scale
length of the halo, respectively.
The $c$-parameter ($c=r_{\rm vir}/r_{\rm s}$, where $r_{\rm vir}$ is the virial
radius of a dark matter halo) and $r_{\rm vir}$ are chosen appropriately
for a given dark halo mass ($M_{\rm dm}$)
by using the $c-M_{\rm h}$ relation
predicted by recent cosmological simulations (Neto et al. 2007).

The bulge of a disk galaxy (for the MW-type)  has a size of $R_{\rm b}$
and a scale-length of $R_{\rm 0, b}$
and is represented by the Hernquist
density profile. The bulge is assumed to have isotropic velocity dispersion
and the radial velocity dispersion is given according to the Jeans equation
for a spherical system.
Although the bulge-mass fraction ($f_{\rm b}=M_{\rm b}/M_{\rm d}$) can be  
a free parameter,
we mainly investigate the MW-type  models 
with $f_{\rm b}=0.17$ and $R_{\rm b}=0.2R_{\rm s}$
(i.e.,  $R_{\rm 0,b}=0.04R_{\rm s}$).

The radial ($R$) and vertical ($Z$) density profiles of the stellar disk are
assumed to be proportional to $\exp (-R/R_{0}) $ with scale
length $R_{0} = 0.2R_{\rm s}$  and to ${\rm sech}^2 (Z/Z_{0})$ with scale
length $Z_{0} = 0.04R_{\rm s}$, respectively.
The gas disk with a size  $R_{\rm g}=2R_{\rm s}$
has the  radial and vertical scale lengths
of $0.2R_{\rm g}$ and $0.02R_{\rm g}$, respectively.
In the present model for the MW-type,  the exponential disk
has $R_{\rm s}=17.5$ kpc and
$R_{\rm g}=35$ kpc.  
In addition to the
rotational velocity caused by the gravitational field of disk,
bulge, and dark halo components, the initial radial and azimuthal
velocity dispersions are assigned to the disc component according to
the epicyclic theory with Toomre's parameter $Q$ = 1.5.
The vertical velocity dispersion at a given radius is set to be 0.5
times as large as the radial velocity dispersion at that point.
%as is consistent with the observed trend of the Milky Way.

Although the present simulation code enables us to investigate
the time evolution of chemical abundances and  dust properties
in a self-consistent manner, we do not investigate 
these properties in the present study. This is firstly because we focus exclusively
on SFHs of disk galaxies (i.e., not chemical evolution),
and secondly because inclusion of such calculations of chemical and dust evolution
is numerically costly and not suitable for a wide parameter study like the present work.
We therefore investigate only dynamical and hydrodynamical evolution of galaxies with
star formation in the present study. 
Initial temperature of gas is set to be $10^4$K for all models.

The total numbers of particles used for dark matter, stellar disk,
gas disk, and bulge in a MW-type disk are
700000, 200000, 100000, and 33400, respectively (i.e., $N=1033400$
is used in total). Since no bulge is included in the M33-type and Dwarf-type models,
the total particle number is 1000000 for the two.
The softening length of dark matter halo ($\epsilon_{\rm dm}$) for each model
is chosen so that $\epsilon_{\rm dm}$ can be the same as the
mean particle separation at the half-mass radius of the halo.
This method is applied  for
determining softening length for stellar particles
($\epsilon_{\rm s}$) in the initial disk of each model.
The softening length is assumed to be the same between old stellar,
gaseous, and new stellar
particles in the present study.
The gravitational softening length for dark ($\epsilon_{\rm dm}$)
and baryonic components ($\epsilon_{\rm s}$)
are 2.1 kpc and 200 pc, respectively, for the MW-type disk.
These values are different in models with different sizes and masses.
In Appendix A,  we discuss how the present results depend on
the resolution of simulations, in particular, the total particle number
used for a disk galaxy, $N_{\rm d}$.

\subsection{Star formation and SN feedback effects}

A gas particle {\it can be} converted
into a new star if (i) the local dynamical time scale is shorter
than the sound crossing time scale (mimicking
the Jeans instability) , (ii) the local velocity
field is identified as being consistent with gravitationally collapsing
(i.e., div {\bf v}$<0$),
and (iii) the local density exceeds a threshold density for star formation ($\rho_{\rm th}$).
The threshold gas density is given in units of the number of hydrogen atoms
per cm$^3$ in the present study just for convenience.
Accordingly, $\rho_{\rm th}=1$ cm$^{-3}$ indeed means 
$\rho_{\rm th}=1 {\rm m}_{\rm H}$ cm$^{-3}$,
where $m_{\rm H}$ is the mass of a hydrogen
atom.
We mainly investigate the models with $\rho_{\rm th}=1$ cm$^{-3}$,
and the dependences of the present results on $\rho_{\rm th}$ 
are briefly discussed in  \S 3.6.

A gas particle can be regarded as a `SF candidate' gas particle
if the above three SF conditions (i)-(iii) are satisfied.
It could be possible to convert some fraction 
of a SF candidate  gas particle
into a new star at each time step until the mass of the gas particle
becomes very small. However, this SF conversion method can increase dramatically
the total number of stellar particles, which becomes  numerically very costly.
We therefore adopt the following SF conversion method that
was already adopted by our previous chemodynamical simulations of galaxies
(Bekki \& Shioya 1998).
A SF candidate $i$-th gas
particle is regarded as having  a SF probability ($P_{\rm sf}$);
\begin{equation}
P_{\rm sf}=1-\exp ( - 
\Delta t {\rho}^{\alpha_{\rm sf}-1} ),
\end{equation}
where $\Delta t$ is the time step width for the gas particle,
$\rho$ is the gas density of the particle,
and $\alpha_{\rm sf}$ is
the power-law slope of the  Kennicutt-Schmidt law
(SFR$\propto \rho_{\rm g}^{\alpha_{\rm sf}}$;  Kennicutt 1998).
A reasonable value of
$\alpha_{\rm sf}=1.5$ is adopted in the present
study.

At each time step   random numbers ($R_{\rm sf}$; $0\le R_{\rm sf}  \le 1$)
are generated and compared with $P_{\rm sf}$.
If $R_{\rm sf} < P_{\rm sf}$, then the gas particle can be converted into
a new stellar one.
In this SF recipe, a gas particle with a higher gas density
and thus a shorter SF timescale ($\propto
\rho/\dot{\rho} \propto \rho^{1-\alpha_{\rm sf}}$)
can be more rapidly converted into a new star owing to the larger
$P_{\rm sf}$. 
We thus consider that the present SF model is a good approximation
for star formation in high-density  gas of disk galaxies.

Each SN is assumed to eject the feedback energy ($E_{\rm sn}$)
of $10^{51}$ erg and 90\% and 10\% of $E_{\rm sn}$ are used for the increase
of thermal energy (`thermal feedback')
and random motion (`kinematic feedback'), respectively.
The thermal energy is used for the `adiabatic expansion phase', where each SN can remain
adiabatic for a timescale of $t_{\rm adi}$.
Although a reasonable value for a single SN explosion
is $t_{\rm adi}=10^5$ yr, 
we adopt $t_{\rm adi}=10^6$ yr in the present study,
because  $t_{\rm adi}$ might be  much longer
for multiple SN explosions in a small local region owing to complicated
interaction between gaseous ejecta from different SNe.  
The energy-ratio of thermal to kinematic feedback is consistent with
previous numerical simulations by Thornton et al. (1998) who investigated
the energy conversion processes of SNe in  detail.
The way to distribute $E_{\rm sn}$ of SNe among neighbor gas particles
is the same as described in Bekki (2013).
The radiative cooling processes
are properly included  by using the cooling curve by
Rosen \& Bregman (1995) for  $100 \le T < 10^4$K
and the MAPPING III code
for $T \ge 10^4$K
(Sutherland \& Dopita 1993).

\begin{figure*}
\psfig{file=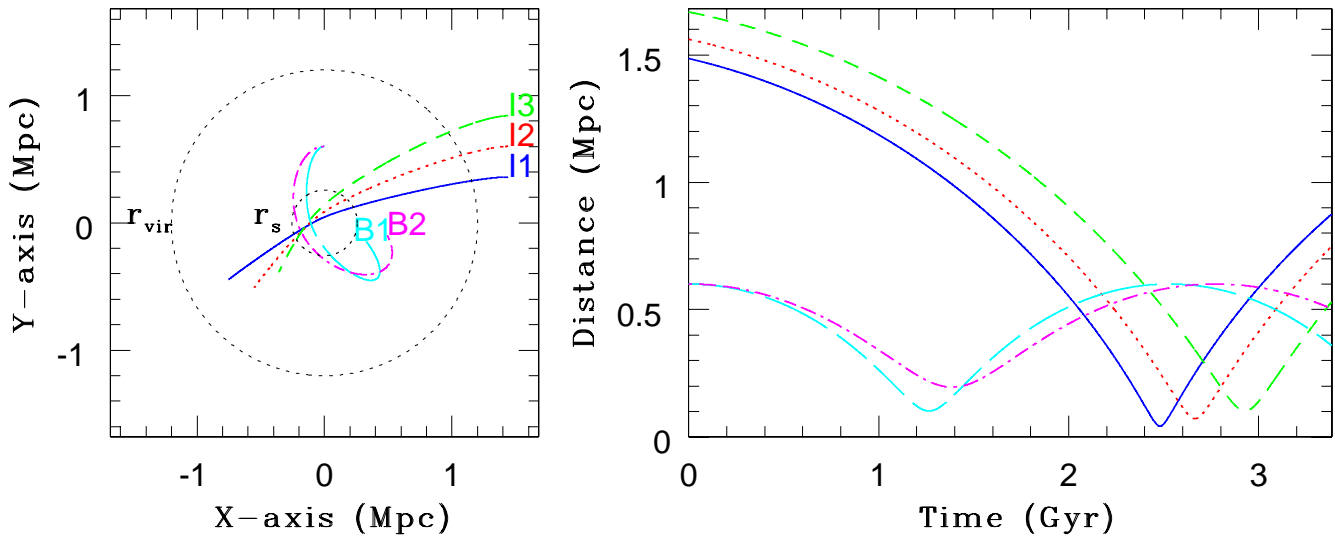,width=18.0cm}
\caption{
The orbit of a disk galaxy projected onto the $x$-$y$ plane (left)
and the time evolution of the galaxy's distance from the cluster center (right)  for the five
orbital types, I1 (blue, solid), I2 (red, dotted), I3 (green, short-dashed),
B1 (cyan, long-dashed), and B2 (magenta, dot-dashed) in the low-mass Virgo
cluster model. The inner and outer circles (black, dotted) indicate the scale radius
($r_{\rm s}$) and the virial radius ($r_{\rm vir}$) of the cluster, respectively.
}
\label{Figure. 1}
\end{figure*}

\subsection{Time-varying ram pressure force}

\subsubsection{A box of ICM}

A disk galaxy is assumed to be embedded in  hot ICM with
temperature  $T_{\rm ICM}$,
density ${\rho}_{\rm ICM}$, and velocity $V_{\rm r}$.
Here $V_{\rm r}$ is the relative velocity of ICM with respect to the velocity
of the disk galaxy. 
The intragroup medium in groups is also referred to as ICM just for convenience
in the present study.
 The strength of ram pressure force on the disk 
is time-dependent and described as follows:
\begin{equation}
P_{\rm ram}(t)=\rho_{\rm ICM}(t) V_{\rm r}^2(t),
\end{equation}
where ${\rho}_{\rm ICM}(t)$ and $V_{\rm r}$(t)
are determined by 3D positions and velocities of a galaxy
at each time step in a simulation  (described later).

In order to avoid huge
particle numbers to represent the entire ICM in clusters
of galaxies (e.g., Abadi et al. 1999),
the ICM is represented by SPH particles 
in a cube with the size of $R_{\rm ICM}$.
A disk galaxy is placed in the exact center of the cube and the direction
of the orbit (within a cluster) is chosen as the $x$-axis  in the Cartesian coordinate
of the cube (Therefore, it should be noted that this $x$-axis is not the same as
the $x$-axis in the orbital calculation of the present study shown in Fig. 1).
The ICM cube is represented by $60^3$  SPH particles each of which
is on the $60 \times 60 \times 60$  cubic lattice.
The initial velocity of each SPH particle for the ICM
is set to be ($V_{\rm r}(t=0)$, 0, 0) for all models
(i.e., the particle flows along the $x$-axis
to the positive $x$ direction).
The ICM  has an uniform distribution
within the cube and $R_{\rm ICM}$ is set to be $6R_{\rm s}$,
and $T_{\rm ICM}$ 
is much higher than the temperature of cold ISM in galaxies.
We include periodic boundary conditions (at $R_{\rm ICM}$)
for the ICM
SPH particles leaving the cube.
In Appendix A,  we briefly discuss how the present results depend on
the total  number of ICM particles ($N_{\rm ICM}$) 
and the box  size ($R_{\rm ICM}$).

The strength of ram pressure of ICM depend on
${\rho}_{\rm ICM}(t)$ and $V_{\rm r}$(t), but it does not depend explicitly on
$T_{\rm ICM}$.
Furthermore,
the present study focuses on the influences of ram pressure on galaxy evolution
in groups and clusters with $M_{\rm h} \ge 10^{13} {\rm M}_{\odot}$ with high ICM temperature
($>10^6$ K).
We therefore assume that 
$T_{\rm ICM}$ is fixed at a certain value for a cluster (no radial temperature
gradient and no temporal variation for simplicity) and determined by 
the initial total halo mass of the cluster ($M_{\rm h}$).
We investigate two cluster models: Low-mass cluster model (`Virgo', labeled as `VI') with 
$M_{\rm h}=10^{14} {\rm M}_{\odot}$ and $T_{\rm ICM}=2.6 \times 10^7$ K
and high-mass cluster model (`Coma',  `CO')  with
$M_{\rm h}=10^{15} {\rm M}_{\odot}$ and $T_{\rm ICM}=1.2 \times 10^8$ K.
These adopted $T_{\rm ICM}$ 
values are consistent with $T_{\rm ICM}$ estimated by $X$-ray observations
for clusters with the above adopted masses  (e.g., Matsumoto et al. 2000).

We also investigate massive  group models with 
$M_{\rm h}=10^{13} {\rm M}_{\odot}$ and $T_{\rm ICM}=5.6 \times 10^6$ K
in order to investigate the influence of the 
halo gas of groups on SFHs of group member galaxies.
These group models are referred to as `Group' (`GR').
The presence of intragroup gas and the influences of the gas on
the evolution of group galaxy evolution have been observationally
confirmed by previous X-ray observational studies for groups
(e.g. Ponman et al. 1996; Helsdon \& Ponman 2003; Rasmussen \& Ponman 2007).
It is therefore worthwhile for the present study to investigate the influences
of ram pressure on galaxy evolution in groups.
The initial total mass  of  ICM  ($M_{\rm ICM}$)
in a cluster/group  is assumed to be $0.15 M_{\rm h}$ in all
models. The adopted value could be large for groups with smaller masses,
because it is unlikely that most intragroup gas is in the form of hot gas.
Furthermore, the radial density profiles of hot gas in the central
regions of groups would  significantly deviate
from those of dark matter.
This means that the present simulation could overestimate the effects
of ram pressure of ICM on galaxies at pericenter passage for groups. 
The basic parameters for these three models are given in Table 2.

\begin{table*}
\centering
\begin{minipage}{180mm}
\caption{Description of the basic parameter values
for the five orbital  models.}
\begin{tabular}{lccccc}
Orbit type/Physical properties
&  I1 & I2 & I3 & B1 & B2 \\
Initial position ($P_{\rm x,0}$,$P_{\rm y,0}$,$P_{\rm z,0}$) &
($1.2 r_{\rm vir}$,$0.3r_{\rm vir}$,0) &
($1.2 r_{\rm vir}$,$0.5r_{\rm vir}$,0) &
($1.2 r_{\rm vir}$,$0.7r_{\rm vir}$,0) &
(0,$0.5r_{\rm vir}$,0) &
(0,$0.5r_{\rm vir}$,0) \\
Initial velocity  ($V_{\rm x,0}$,$V_{\rm y,0}$,$V_{\rm z,0}$) &
($-0.3 v_{\rm c}$,0,0) &
($-0.3 v_{\rm c}$,0,0) &
($-0.3 v_{\rm c}$,0,0) &
(0,$-0.3 v_{\rm c}$,0,0) &
(0,$-0.5 v_{\rm c}$,0,0) \\
\end{tabular}
\end{minipage}
\end{table*}

\subsubsection{Galaxy orbits within clusters}

A cluster of galaxies is assumed to have the NFW density profile and the $c$-parameter
and $r_{\rm vir}$ are determined by $M_{\rm h}$ by using the formula by Neto et al. (2007).
The orbit  of a disk galaxy is determined by the adopted density profile of the cluster and 
the initial orbital plane is set to be the same as the $x$-$y$ plane in the cluster.
The initial 3D position  and velocity  of the galaxy are denoted as
($P_{\rm x,0}$, $P_{\rm y,0}$, $P_{\rm z,0}$) 
and ($V_{\rm x},0$, $V_{\rm y,0}$, $V_{\rm z,0}$), respectively.
The density of ICM surrounding the galaxy 
at each time step is simply the dark matter density at the galaxy position
multiplied by 0.15 (i.e., ICM mass fraction). Using the 3D velocity components
of the galaxy at each time step, $V_{\rm r}(t)$ can be easily calculated. 

We mainly investigate three `infall' models (I1, I2, and I3) and two `bound' ones
(B1 and B2) for the orbital evolution of disk galaxies in clusters. For the infall models,
$P_{\rm x,0}=1.2r_{\rm vir}$, 
$P_{\rm y,0}=f_{\rm x}r_{\rm vir}$ (where $f_{\rm x}$ is a parameter that corresponds
to an impact parameter and thus controls the pericenter distance),
$P_{\rm z,0}=0$,
$V_{\rm x,0}=-f_{\rm v} v_{\rm c}$ (where $v_{\rm c}$ is the circular velocity at
the initial galaxy position and $f_{\rm v}$ is fixed at 0.3),
$V_{\rm y,0}=0$,
and $V_{\rm z,0}=0$. 
Accordingly, a galaxy can infall onto a cluster environment from outside the cluster
virial radius and then pass through the cluster core. 
The smaller $f_{\rm x}$ is,
the smaller the orbital pericenter distance is in these infall models.
The $f_{\rm x}$ values are 0.3, 0.5, and 0.7 for I1, I2, and I3 models, respectively.

For the bound models, 
$P_{\rm x,0}=0$, 
$P_{\rm y,0}=f_{\rm x}r_{\rm vir}$,
$P_{\rm z,0}=0$, 
$V_{\rm x,0}=-f_{\rm v} v_{\rm c}$,
$V_{\rm y,0}=0$
and $V_{\rm z,0}=0$. 
The models with smaller $f_{\rm v}$ show more eccentric orbits within
a cluster for a given $f_{\rm x}$.
We mainly show the results of the models with $f_{\rm x}=0.5$ and 
$f_{\rm v}=0.3$ (B1) and $f_{\rm v}=0.5$ (B2) in the present study.
The initial 3D positions and velocities for the five models
are given in Table 3.
The orbital evolution of the five models (I1, I2, I3, B1, and B2) within
a cluster with $M_{\rm h}=10^{14} {\rm M}_{\odot}$ is shown in
Fig. 1.

\subsubsection{Disk inclination}

The inclination angle of a disk with respect to the direction of the orbit in
a cluster can be a key parameter that controls the total amount of gas stripped
from the disk and thus the SFH. Since the orbital plane of a disk is the same as the
$x$-$y$ plane, the spin axis of a disk
with respect to the $x$-$y$ plane is a key parameter in the present study.
The spin of the disk  galaxy
is specified by two angles $\theta$ and
$\phi$ (in units of degrees), where
$\theta$ is the angle between
the $z$-axis and the vector of the angular momentum of the disk,
and $\phi$ is the azimuthal angle measured from $x$ axis to
the projection of the angular momentum vector of the disk onto
the $x$-$y$ plane.
We mainly investigate the following three models: 
`face-on' (labeled as `FA') with  $\theta=90$ and $\phi=0$,
`edge-on' (`ED') with  $\theta=0$ and $\phi=0$,
and `inclined'  (`IN') with  $\theta=45$ and $\phi=30$.
In the face-on inclination model,  a disk can be more strongly influenced
by ram pressure stripping. 

\begin{figure*}
\psfig{file=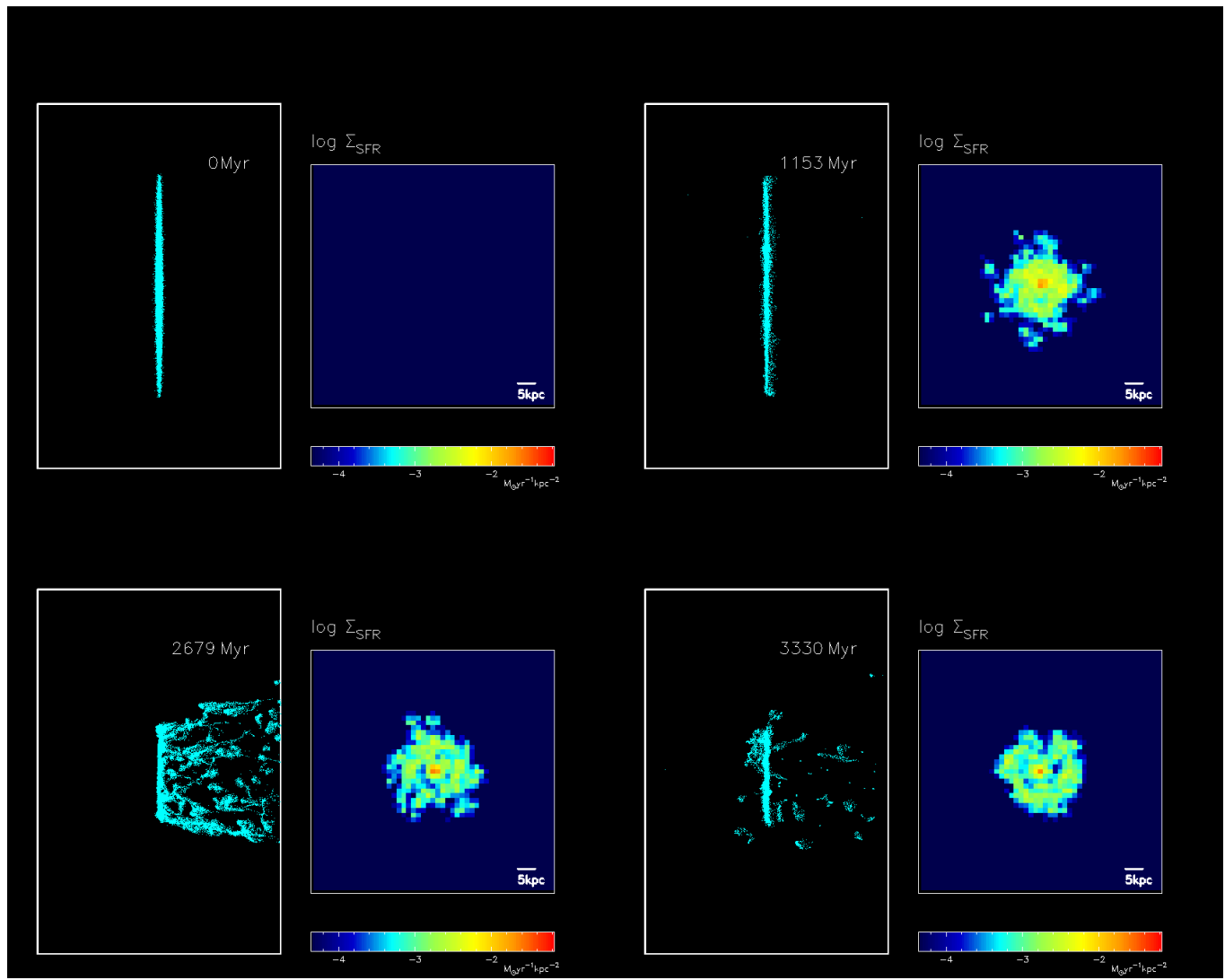,width=18.0cm}
\caption{
The gas distribution (edge-on view) of the disk galaxy
(left,  rectangular box) and the 2D distribution for the
SFR density of young stars (right, square box)
for the fiducial model with the MW-type disk, the Virgo cluster model,
I1 orbit, and face-on disk inclination (labeled as `VI-I1-MW-IN') at four
different time steps.
The rectangular  panel 
for the projected gas distribution shows the time in units of Myr at the upper
right corner.
The 2D SFR density ($\Sigma_{\rm SFR}$ in logarithmic scale) 
 is given in units of ${\rm M}_{\odot}$ yr$^{-1}$ kpc$^{-2}$,
and the thick bar in the lower right corner shows a physical scale of 5 kpc.
}
\label{Figure. 2}
\end{figure*}

\subsection{Parameter study}

We mainly investigate the SFHs and 2D distributions
of SF regions  in MW-type, M33-type, and Dwarf-type disk galaxies
for low-mass (Virgo) and high-mass (Coma) cluster models. 
For comparison,
we also run an `isolated model' in which ram pressure is not included at all.
Since gas infall onto disks is not included in this isolated model,
star formation can gradually decline in this model (i.e., this model does not
correspond to field disk galaxy populations for which continuous gas infall might
be possible for fueling SF).

We first  show the results of the `fiducial' model
and then describe the parameter dependences of the present numerical simulations.
In the fiducial model,  the MW-type disk (`MW'),  the Virgo cluster model (`VI'), face-on
inclination (`FA'), and I1 orbit (`I1')
are adopted. The key parameters for the fiducial model are given in Table 4.
This fiducial model is also labeled as `VI-I1-MW-FA' and
each model in the present study is labeled like this.
For example, `CO-B1-DW-IN' model means that a dwarf-type disk galaxy (`DW')
with a gas disk inclined with respect to its orbital direction
(`IN') is moving in a Coma-like cluster (`CO')  with
a highly eccentric bound orbit (`B1').

We mainly investigate SFHs and H$\alpha$-to-optical-disk-size ratio ($s_{\rm sf}$) 
in disk galaxies under strong ram pressure. We define $s_{\rm sf}$ as follows:
\begin{equation}
s_{\rm sf}=\frac{ R_{\rm sf} }{ R_{\rm s} },
\end{equation}
where $R_{\rm s}$ is the size of the initial stellar disk composed only of
old stars and $R_{\rm sf}$ is the size of the star-forming disk  within which 98\% of
all new stars (with ages less than the lifetime of a $8m_{\odot}$ star) are included.
Although we could define $R_{\rm sf}$ as the  distance 
of the most distant new  star from its host galaxy's  center, 
we do not define $s_{\rm sf}$ as such. This is mainly because 
$s_{\rm sf}$ can change dramatically even within consecutive two time steps for data output,
if $R_{\rm sf}$ is the  distance of the most distant new star from the galaxy center.
 Therefore,  the number of 98\% has no physical meaning and $s_{\rm sf}$
is used just for roughly estimating the size of the simulated H$\alpha$ disk 
in a model and comparing it with the observed H$\alpha$ disk.
The half-mass radius of new stars in a simulated disk ($R_{\rm h, sf}$),
which could be more physically meaningful, 
is also investigated. This $R_{\rm h, ns}$, however,  does not change so much
in comparison with $R_{\rm sf}$ in the present models.

We also investigate 2D densities of SFRs and H$\alpha$ by using the following
formula (Kennicutt 1998);
\begin{equation}
{\rm SFR}  ({\rm M}_{\odot} {\rm yr}^{-1})=
\frac { L({\rm H}\alpha)  } 
{ 1.26 \times 10^{41} {\rm erg} {\rm s}^{-1} }.
\end{equation}
First we divide the simulated disk into $50 \times 50$ cells
and estimate the SFR density ($\Sigma_{\rm SFR}$ [${\rm M}_{\odot}$ yr$^{-1}$ kpc$^{-2}$])
for each cell.
Then H$\alpha$ luminosity density for each cell can be estimated by using
the above formula and $\Sigma_{\rm SFR}$.

\begin{figure*}
\psfig{file=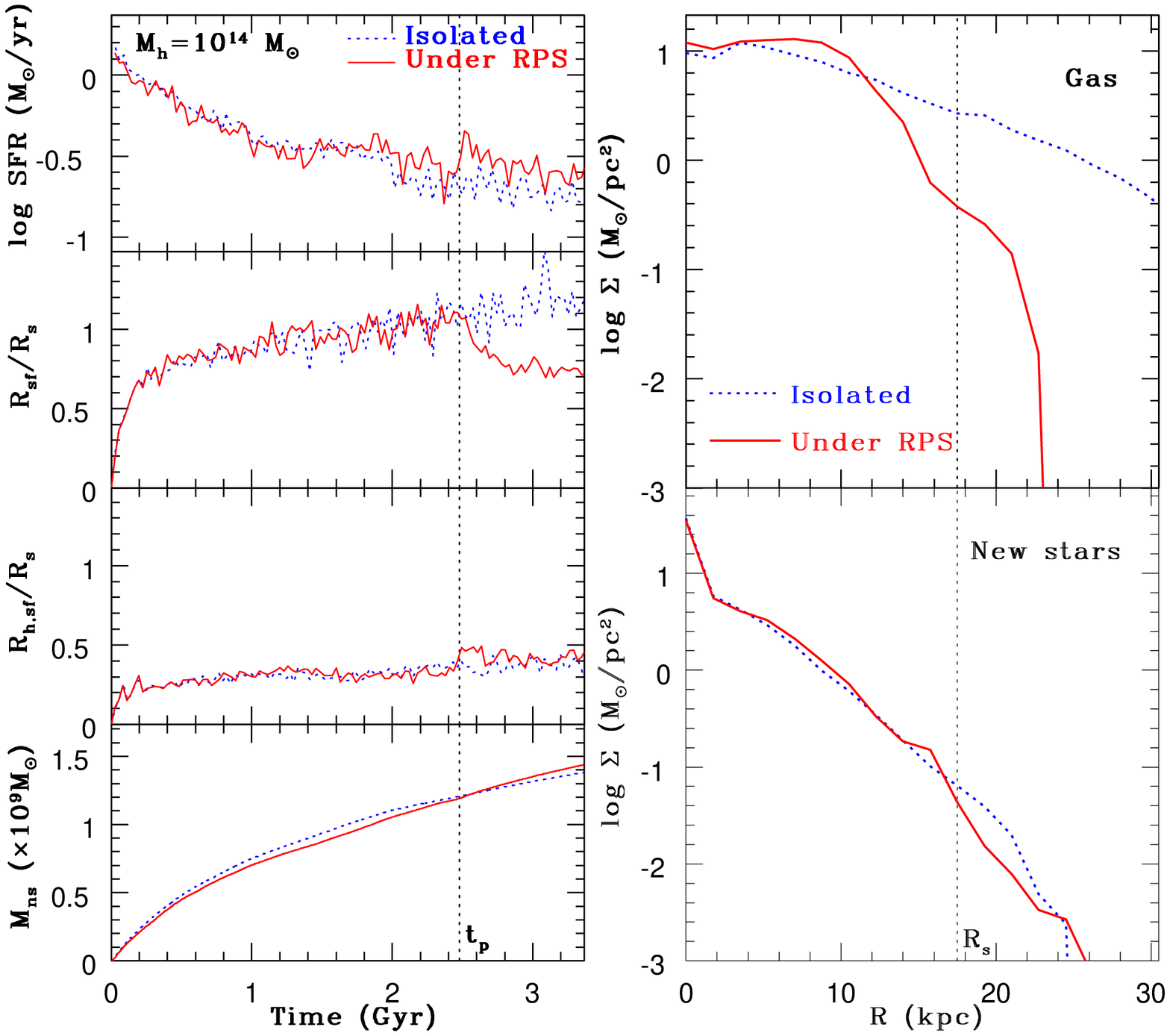,width=15.0cm}
\caption{
 Left panel: The time evolution of star formation rate (SFR, top), 
the ratio of star-forming disk to initial stellar disk ($R_{\rm sf}/R_{\rm s}$, 
second from the top),
the half-mass radius of new stars normalized by initial stellar disk size
($R_{\rm h, sf}/R_{\rm s}$, second from the bottom),
and the total mass of new stars ($M_{\rm ns}$, bottom)
for the isolated model (blue, dotted)
and the fiducial model under ram pressure stripping (labeled as `RPS')
within the Virgo-like cluster with $M_{\rm h}=10^{14} {\rm M}_{\odot}$
(red, solid). 
The black dotted line
indicates the time of pericenter passage ($t_{\rm p}$).
Right panel: The final projected distributions of gas (upper) and new stars
(lower) for the isolated model (blue, dotted)
and for the fiducial model (red, solid).
}
\label{Figure. 3}
\end{figure*}

As previous numerical studies suggested (e.g., Kronberger et al. 2008; Bekki et al. 2010),
high pressure of ICM can have positive (SF enhancement) and negative (SF reduction)
effects on SFHs of disk galaxies in clusters.
In order to understand  which of the two effects can be more dominant for each model,
we investigate the ratio of the total mass of new stars formed in a disk under
ran pressure ($M_{\rm ns}$) to that in an isolated disk ($M_{\rm ns, 0}$).
The ratio is accordingly represented by $\epsilon_{\rm sf}$ as follows:
\begin{equation}
\epsilon_{\rm sf}=\frac{ M_{\rm ns} }{ M_{\rm ns,0} }.
\end{equation}
If this $\epsilon_{\rm sf}$ is larger than 1 in a model,
it means that gas is more efficiently
converted into new stars during  $3.4$ Gyr evolution of the disk within a group/cluster
in comparison with the isolated model
(i.e., ram pressure can enhance the conversion of cold gas into new stars). It should be
noted here that the isolated model does not correspond to field galaxies (for which
gas might  continue to be accreted onto the gas disks so that SFRs can be 
kept constantly high,
unlike the isolated model of the present study). Therefore,
$\epsilon >1 $ in a model does not mean that global SF in the disk 
is enhanced by ram pressure in comparison with field galaxies.

\begin{figure*}
\psfig{file=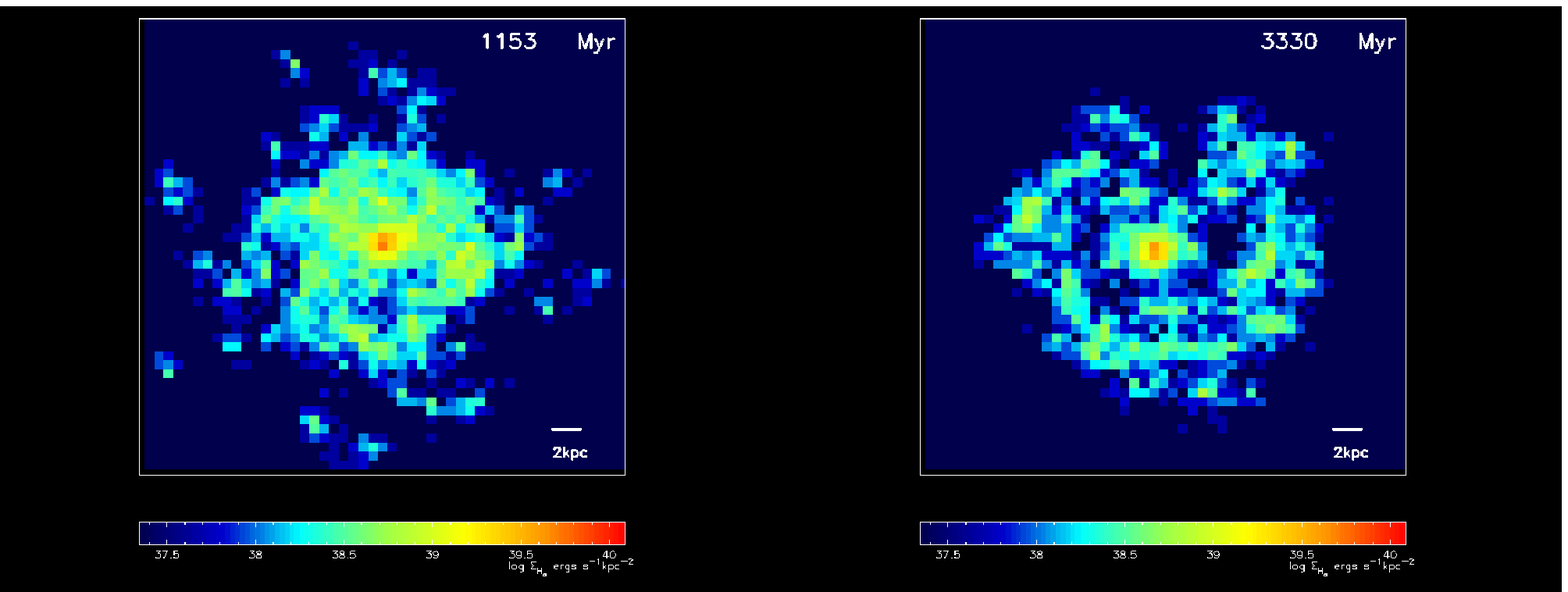,width=18.0cm}
\caption{
The 2D distributions of H$\alpha$ emission densities 
($\Sigma_{\rm H_{\alpha} }$, in logarithmic scale) for the fiducial model
at $T=1153$ Myr (left) and $T=3330$ Myr (right). These 2D H$\alpha$ maps are
from 2D SFR density maps shown in Fig. 2. 
Clearly, the disk at $T=3330$ Myr (i.e., after ram pressure stripping)
shows an outer ring-like structure in the H$\alpha$ map.
}
\label{Figure. 4}
\end{figure*}

\begin{table}
\centering
\begin{minipage}{80mm}
\caption{Description of the basic parameter values
for the fiducial  model.}
\begin{tabular}{lc}
Physical properties &  Adopted values \\
Cluster model & Low-mass (Virgo, $M_{\rm h}=10^{14} {\rm M}_{\odot}$)  \\
Galaxy-type  & MW  ($M_{\rm d}=5.4 \times 10^{10} {\rm M}_{\odot}$)\\
Orbit  &  I1 \\
Disk inclination  & ($\theta$,$\phi$)=(90$^{\circ}$,0$^{\circ}$) \\
\end{tabular}
\end{minipage}
\end{table}

\begin{figure*}
\psfig{file=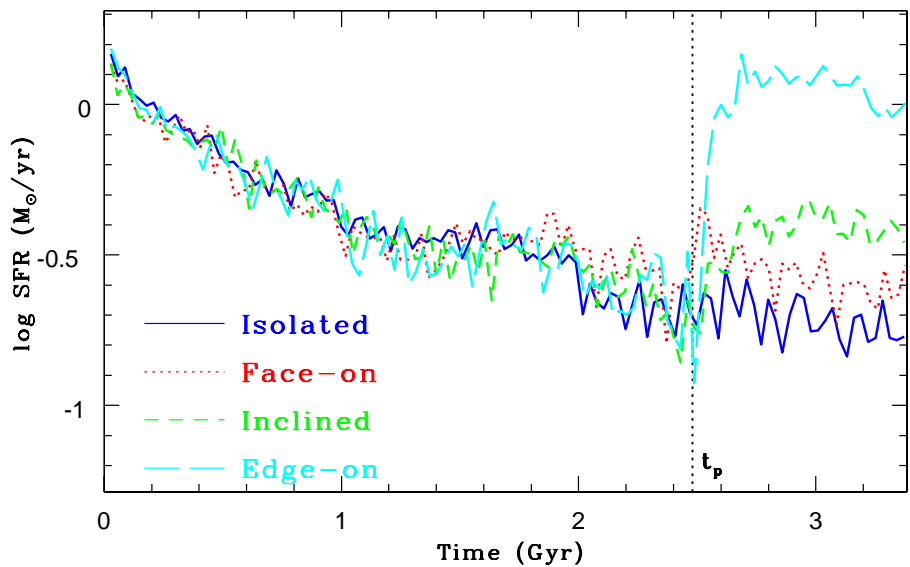,width=12.0cm}
\caption{
The time evolution of  SFRs in the MW-type disk galaxies with 
Face-one (red, dotted), Inclined (green short-dashed), and Edge-on (cyan,
long-dashed) disk orientation models within the low-mass Virgo cluster. 
For comparison, the isolated disk model is shown by a blue solid  line.
The time at the pericenter passage  is indicated by a black dotted line
and marked with `$t_{\rm p}$'.
It is clear that the Edge-on model shows a greater enhancement of SF after
pericenter passage owing to the stronger compression of ISM by ram pressure.
}
\label{Figure. 5}
\end{figure*}

\begin{figure*}
\psfig{file=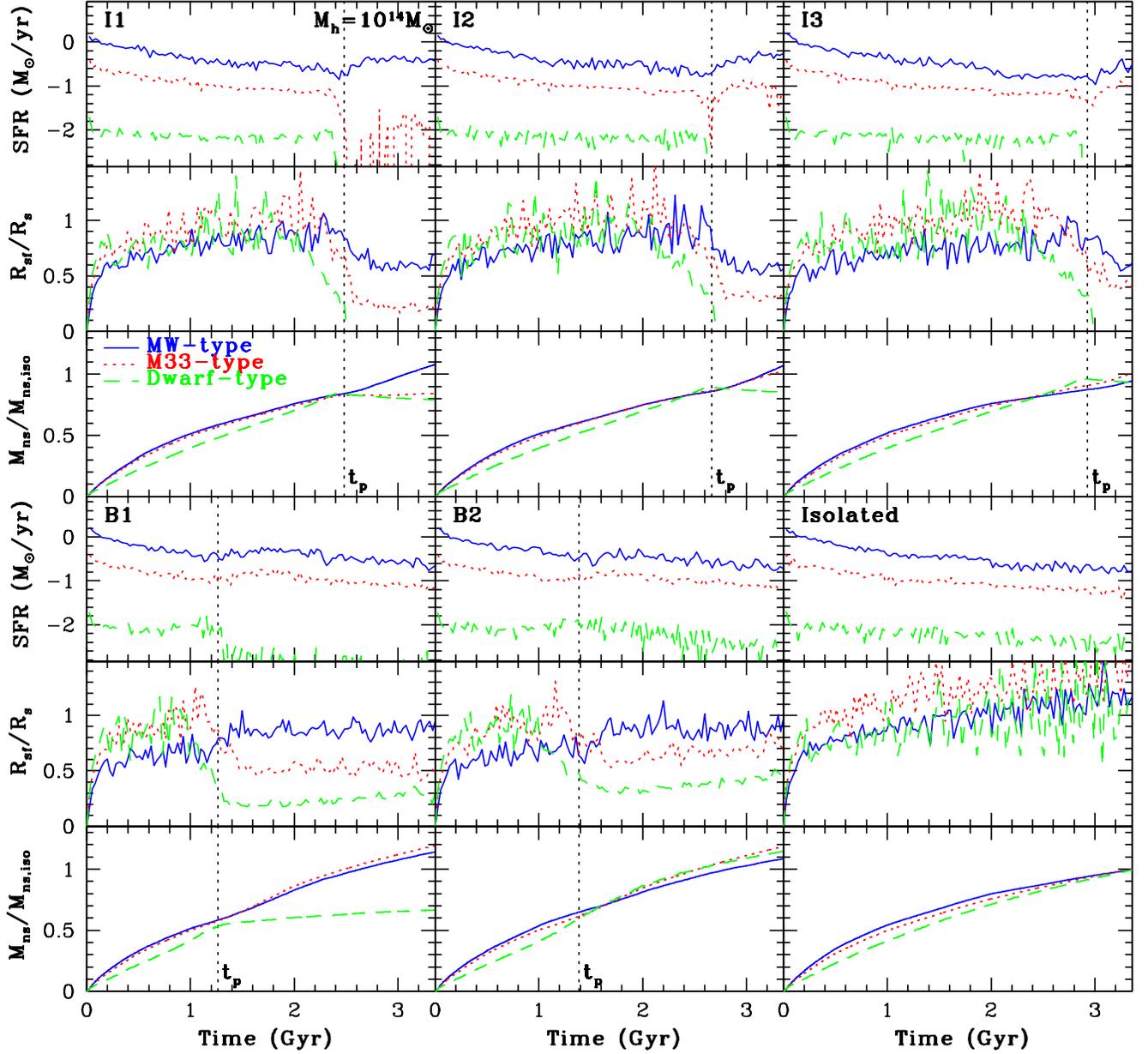,width=18.0cm}
\caption{
The time evolution of SFRs (in logarithmic scale),  $s_{\rm sf}$ ($=R_{\rm sf}/R_{\rm s}$),
and $\epsilon_{\rm sf}$ ($=M_{\rm ns}/M_{\rm ns,0}$) for 15 different disk galaxies
with different orbits and halo masses in the low-mass Virgo cluster model
with $M_{\rm h}=10^{14} {\rm M}_{\odot}$. Each of six sets of three frames shows  
the results of the three different disk galaxies, MW-type (blue, solid),
M33-type (red, dotted), and Dwarf-type (green, dashed), with a specific orbital type
(I1, I2, I3, B1, and B2, indicated  in the upper left corner of the top panel).
For comparison, the results of the three isolated models are also shown. 
The time of pericenter passage ($t_{\rm p}$) is shown by a vertical black dotted line
in each panel.
}
\label{Figure. 6}
\end{figure*}

\begin{figure*}
\psfig{file=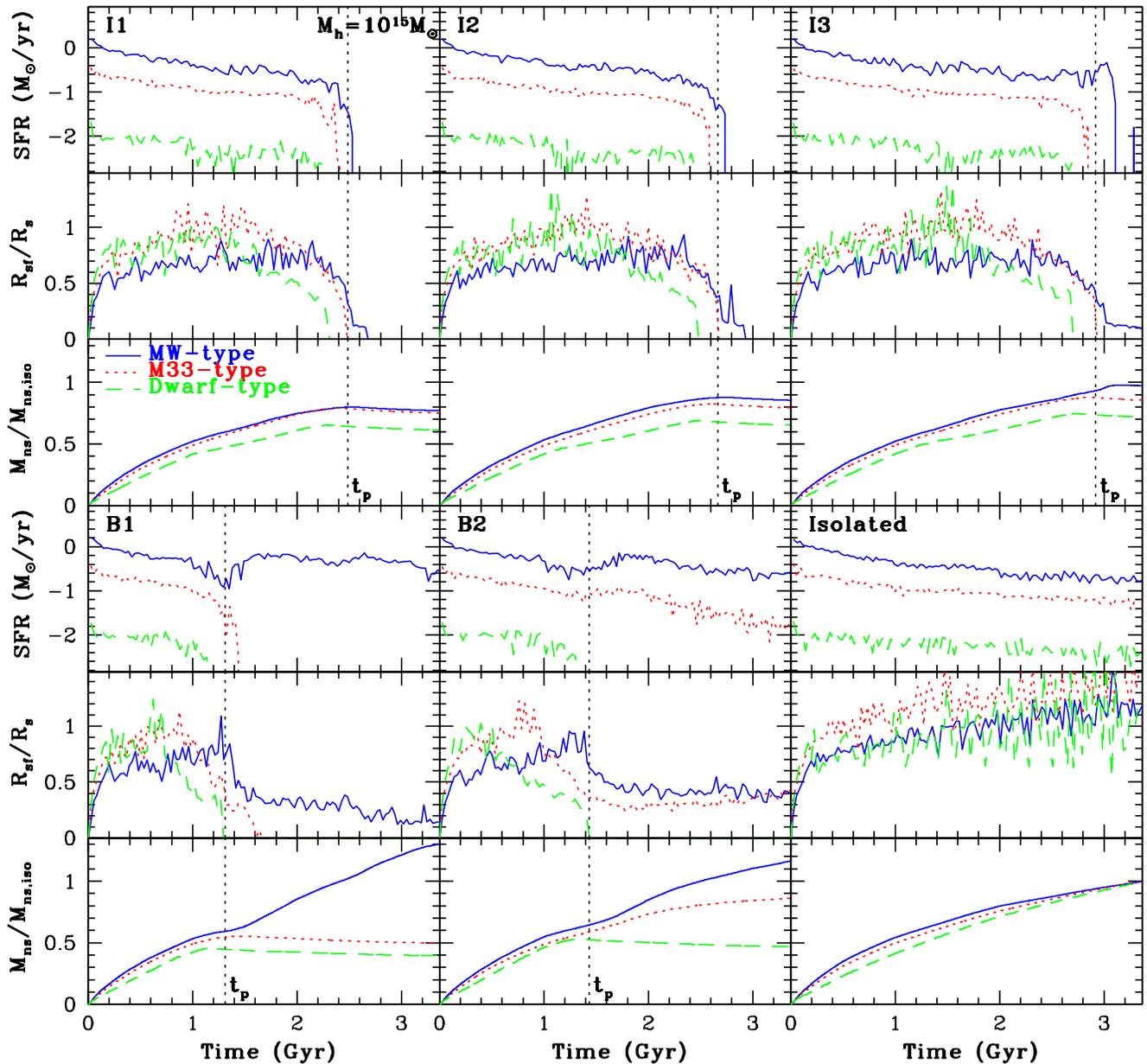,width=18.0cm}
\caption{
The same as Fig. 6 but for the Coma model with $M_{\rm h}=10^{15} {\rm M}_{\odot}$.
}
\label{Figure. 7}
\end{figure*}

\begin{figure*}
\psfig{file=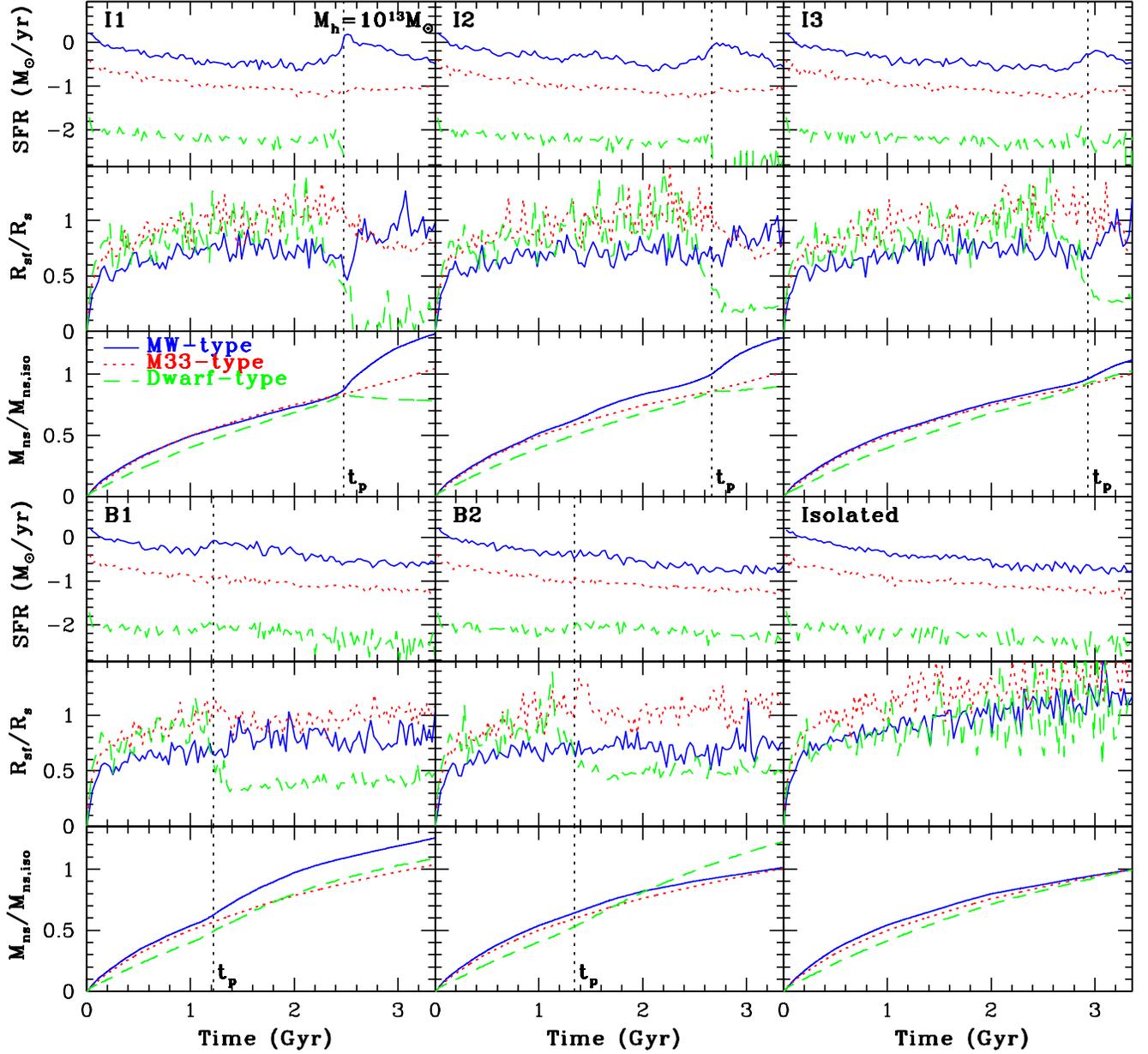,width=18.0cm}
\caption{
The same as Fig. 6 but for the Group model with $M_{\rm h}=10^{13} {\rm M}_{\odot}$.
}
\label{Figure. 8}
\end{figure*}

\begin{figure*}
\psfig{file=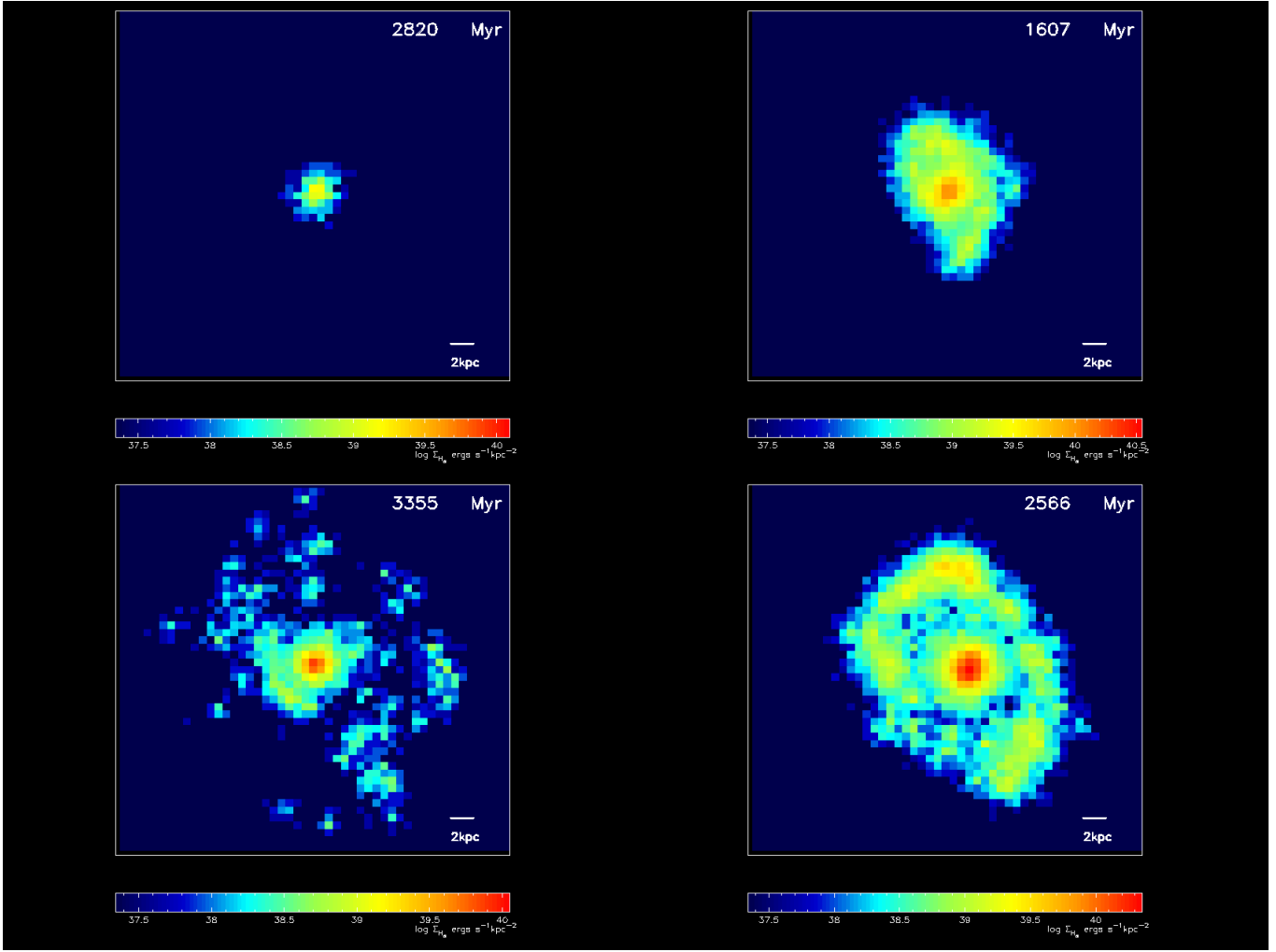,width=18.0cm}
\caption{
A collection of four characteristic H$\alpha$ morphologies in disk galaxies
under ram pressure derived from four models, CO-I2-MW-IN (upper left),
CO-B1-MW-IN (upper right), VI-B1-MW-IN (lower left), and GR-I1-MW-IN (lower right). 
 These H$\alpha$ maps
show very compact and truncated (upper left), one-sided or crescent-like (upper right),
outer (broken), ring-like (lower left), and outer high-density ring-like distributions
(lower right). These four are chosen from many H$\alpha$ distributions of models at
different time steps.
}
\label{Figure. 9}
\end{figure*}

\begin{figure*}
\psfig{file=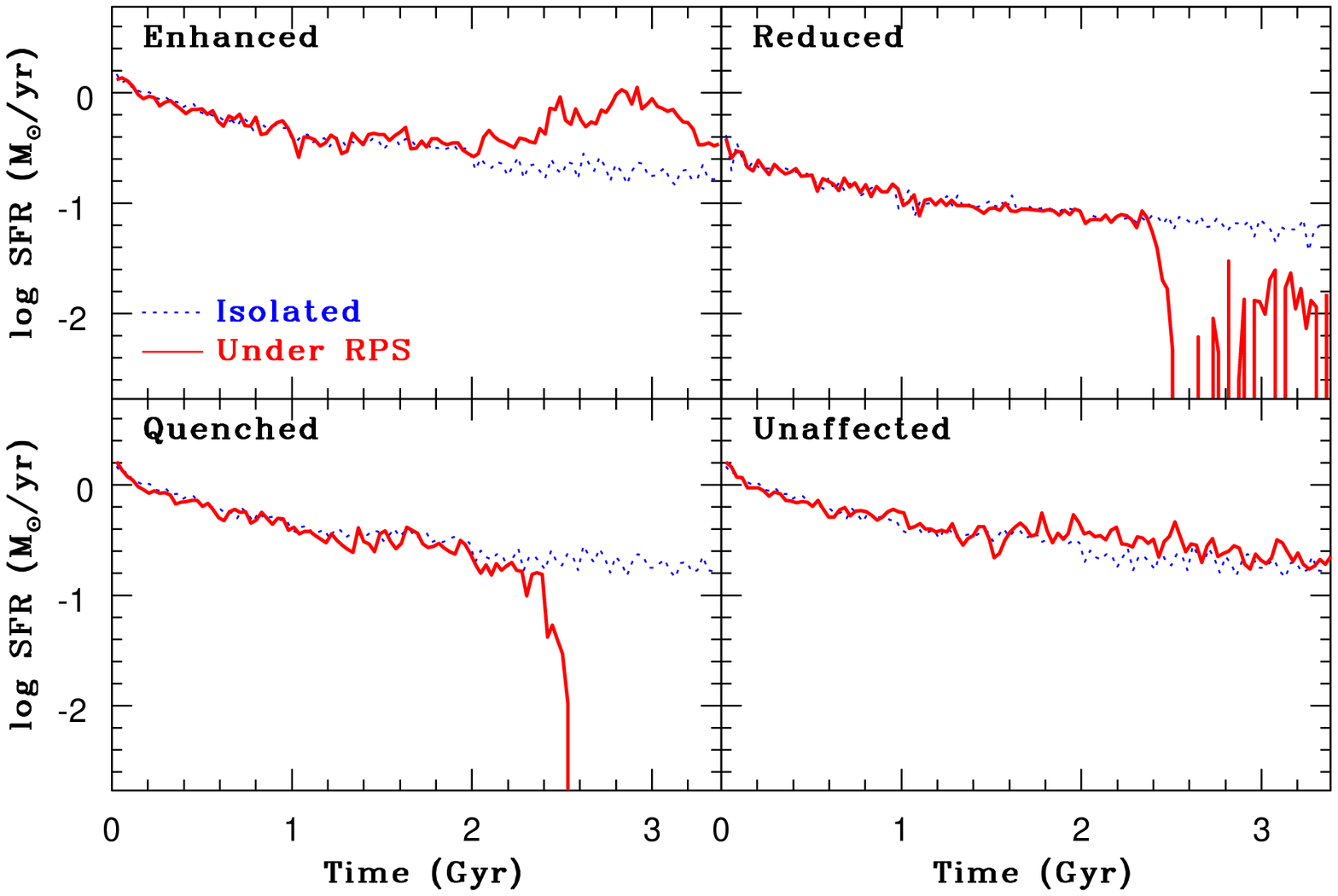,width=18.0cm}
\caption{
The time evolution of SFRs for four representative SF modes derived for disk galaxies
under ram pressure
(red, solid): 
 Enhanced (upper left, the model CO-I2-MW-ED),
Reduced (upper right, VI-I1-M33-IN),
Quenched (lower left, CO-I1-MW-IN), and
Unaffected (lower right, VI-B2-MW-IN).
For comparison, the isolated model is shown by a blue dotted
line in each panel. Ram pressure does not always reduce or quench galactic star formation,
but it can enhance significantly depending on galaxy masses, inclination angles, pericenter
distances, and groups/cluster masses.
The `Unaffected' SF mode means that SF of galaxies are much less influenced
by ram pressure: It does not mean that SF is not influenced by ram pressure
at all (very slight increase of SF can be seen in the model shown in the
lower right panel).
}
\label{Figure. 10}
\end{figure*}

\section{Results}

\subsection{The fiducial model}

Fig. 2 shows how the 2D distributions of cold gas and SFR in a MW-type disk
galaxy changes with time when the galaxy passes  through the central core of 
a Virgo-like cluster. In the early phase of its orbital evolution within
the low-mass cluster,  the ram pressure of ICM is rather weak and therefore can   
influence only the very outer part of the gas disk ($T=1.2$ Gyr). 
Consequently, new stars can continue to form throughout the gas disk as efficiently
as the isolated disk in this early evolution phase.
After the pericenter passage ($r_{\rm p}=42$ kpc at $T=2.5$ Gyr),
the gas disk can be quickly and efficiently stripped by ram pressure so that
numerous small gas clumps can form behind the disk ($T=2.7$ Gyr). 
These small clumps have head-tail structures and most of them can be completely
stripped from the disk galaxy (i.e., outside $r_{\rm vir}$) during the  pericenter passage.
However, a minor fraction of once stripped gas clumps can be returned back to the 
gas disk when the galaxy goes away from the orbital pericenter,
because ram pressure can become weak again ($T=3.3$ Gyr). These clumps falling back to
the original gas disk can be observed as high-velocity clouds in the galaxy.

The details of these stripping processes can be seen in the attached animation,
for which the color codes are exactly the same as those in Fig. 2.
The attached animation (ram.f1.avi for high-resolution
and ram.f1.mov for low-resolution) clearly shows (i) how the gas disk of the disk
galaxy in the fiducial model is stripped by ram pressure of ICM and
(ii) how the 2D distribution of the star-forming regions accordingly changes
during the disk evolution. 
This animation enables  readers to understand better
the following points that are not clearly shown in Fig. 2:
(i) the formation processes of  numerous small gas clumps
during ram pressure stripping,   (ii) the formation head-tail structures of the clumps,
(iii) the re-accretion process of the clumps, and (iv)
the sudden truncation of SF in the outer
disk during ram pressure stripping.

Owing to the severe truncation of the outer gas disk by ram pressure stripping,
SF in the outer disk ($R \sim R_{\rm s}$) can be completely
shut down. As a result of this, the 2D $\Sigma_{\rm SFR}$ distribution shows
a sharp edge around $R \sim 12$ kpc at  $T=3.3$ Gyr.
It is intriguing that the outer 2D $\Sigma_{\rm SFR}$ distribution looks like a ring
with a central concentration. This ring-like distribution of new stars is characteristic
for disk galaxies under strong ram pressure stripping, as discussed later.
The left panel of 
Fig. 3 shows that in spite of efficient gas stripping,
the SFR of the disk galaxy can become higher than the SFR of the isolated model
at $T>2$ Gyr. 
A physical explanation for this moderate  SF enhancement is as follows.
When the disk galaxy becomes closer to the orbital pericenter,  ram pressure becomes
moderately strong and starts to compress the inner part of the gas disk
so that a significant fraction of gas particles can have higher gas densities.
This compression of gas by ram pressure can be the strongest at the pericenter passage
of the galaxy. As a result of this,  SF can be moderately  enhanced  largely
in the central region of the disk at the pericenter passage.
The physical mechanism of this is discussed more extensively in
\S 3.5.

The left panel of
Fig. 3 shows that $s_{\rm sf}$ ($=R_{\rm sf}/R_{\rm s}$ corresponding to
the observed  H$\alpha$-to-optical-disk-size ratio) can become significantly smaller
after the pericenter passage owing to the truncated SF in the outer disk. 
The half-mass radius of new stars ($R_{\rm h, ns}$), however, does not
change significantly during ram pressure stripping.
This means that ram pressure can only influence the outer star-forming
disk in this model. 
The total mass of new stars ($M_{\rm ns}$) formed during $\sim 3.4$  Gyr evolution 
of the disk is only slightly larger than that in the isolated model, which means
that the positive effect of ram pressure on SF (i.e., SF enhancement)
can surpass the negative one (i.e., SF reduction) in this MW-type disk galaxy.
It should be noted, however, that it depends on $M_{\rm h}$,  $r_{\rm p}$, and $\theta$
($\phi$)
of disk galaxies
on  which of these two effects
can be more dominant in SFHs of disk galaxies for a given environment
(This point is later discussed in \S 3.2.)

The right two panels of Fig. 3 show that the final projected 
density distributions ($\Sigma$)
of gas are quite different between the isolated and fiducial models, though
the distributions of new stars are not so different between the two.
The gaseous distributions with almost constant $\Sigma$ for 4 kpc $\le R\le$ 10 kpc
and very steep $\Sigma$ decline beyond $R\sim 10$ kpc 
in the fiducial model clearly mean that ram pressure can significantly
increase the gas density in the inner region of the gas disk. The inner region
with constant $\Sigma$ is roughly coincident with the location of
a ring-like structure of  young stars shown in Fig. 2 (and later in Fig. 4).
This coincidence provides the following explanation for the origin of the
ring-like structure. During efficient ram pressure stripping of the outer
gas disk (which ends up with the sharp $\Sigma$ decline at $R>$ 10 kpc in Fig.3), 
the inner gas disk can be strongly compressed by ram pressure. 
Consequently,  the gas density
of the inner gas disk (4 kpc $\le R \le$ 10 kpc) can become so
high that new stars can form from the gas disk.
The gas at the  circumnuclear  region ($R\sim $ 2 kpc) can be rapidly consumed
by star formation so that the SFR there can be relatively 
low at the time of pericenter
passage. Thus, a ring-like young star-forming region can be formed 
after ram pressure stripping.

The right panel of Fig. 4 shows that (i) SF can be actively ongoing in the central bulge 
region  and (ii) the disk has an intriguing ring-like H$\alpha$ morphology 
after ram pressure stripping of the outer gas disk ($T=3.3$ Gyr). 
The outer sharp edge of the H$\alpha$ ring corresponds  to the outer edge of the truncated
gas disk. Given that such a ring-like structure can not be seen in the disk at $T=1.2$ Gyr
(left panel  of Fig. 4) and in the isolated model,
the formation of the ring is caused by the effects of ram pressure on the gas disk.
Such a ring-like distribution of star-forming regions is indeed observed for NGC 4569 
in the Virgo cluster (e.g., Koopmann \& Kenny 2004).
The simulated ring-like H$\alpha$ morphology is one of characteristic distributions
of H$\alpha$ in disk galaxies under strong ram pressure, as described later.

\subsection{Parameter dependence}

Figs. $5-8$ summarize the dependences of the time evolution of SFRs,
$s_{\rm sf}$ (corresponding to H$\alpha$-to-optical-disk-size ratio), and
$\epsilon_{\rm sf}$ (long-term conversion efficiencies of cold ISM into new stars)
on disk inclination angles ($\theta$ and $\phi$),  halo masses of galaxies ($M_{\rm h}$),
orbits (e.g., $r_{\rm p}$), and group/cluster masses ($M_{\rm h}$). 
For Figs. $5-7$, the results of the disk models with $\theta=45^{\circ}$
and $\phi=30^{\circ}$ (`IN' models) are shown,  because such inclined disk configurations
with respect to the orbits of galaxies
can be more typical and more realistic for real ISM-ICM interaction  in groups and clusters.

\subsubsection{Disk inclination}

As shown in Fig. 5,  SFHs of MW-type galaxies after pericenter passage
depend strongly on disk inclination angles for the low-mass Virgo cluster model.
Clearly, SF can be  more strongly enhanced by ram pressure in the edge-on disk inclination
than in the face-on and inclined ones. Ram pressure can be responsible both for
efficient gas stripping and for strong compression of gas. For the edge-on model,
the latter effect is more dominant than the former one so that SF can be significantly
enhanced in the central region of the gas disk. The level of SF enhancement
in the edge-on model is significant (i.e., similar to the initial active SF phase)
and can continue for $\sim 1$ Gyr after pericenter
passage. Either  `reactivation' or `rejuvenation' would be a better word for describing the
effect of ram pressure on the galactic SFH of  the edge-one model.

\subsubsection{Galaxy mass}

As shown in Fig. 6,  both SFRs and $s_{\rm sf}$  can be more significantly
reduced in less massive disk galaxies for the I1 orbit in the Virgo cluster
model with $M_{\rm h}=10^{14} {\rm M}_{\odot}$.
The MW-type galaxy shows a slightly enhanced SFR after pericenter 
passage whereas the M33-type  shows a  sudden and significant SFR decrease
in the I1 orbit.  
The $s_{\rm sf}$ parameter can be reduced by $\sim 40$\% 
for the MW-type and by $\sim 80$\% for the M33-type in this orbit,
which means that the M33-type has a more compact H$\alpha$ disk.
SF in the Dwarf-type 
can be suddenly  shut down (`SF quenching')
and thus $s_{\rm sf}$ becomes 0 after pericenter-passage
owing to the complete stripping of cold gas from the disk by ram pressure
in this I1 orbit. 
The $\epsilon_{\rm sf}$ parameter 
can become larger than 1 for the MW-type owing to the moderate SF enhancement
in the central bulge region whereas it can be significantly lower than 1
for M33-type and Dwarf-type in this  I1 orbit.
These results suggest that SFHs and 2D distributions of SF regions 
in less massive disk galaxies can be more 
strongly influenced by ram pressure in clusters.

The modest SF enhancement of the MW-type and the sudden SF quenching
of the Dwarf-type derived in the I1 orbit
can be seen in other orbits of the Virgo model and in some orbits of the Coma
and the Group models. However, the significant SF reduction of the M33-type
derived in the I1 orbit can not be clearly seen in other orbits of the Virgo
models. This simply means that SFHs of disk galaxies can be determined not only
by their masses but also by other parameters  ($r_{\rm p}$ and group/cluster masses).
The dependences of the time evolution of $s_{\rm sf}$
on galaxy masses derived in the I1 orbit of the Virgo  model
can be seen in other orbital types (I2, I3, B1, and B2)
of the Virgo model and some orbits of the  Coma cluster model (See Fig. 7).
The degrees  of 
SF and $s_{\rm sf}$ reduction by ram pressure, however, can depend on
orbital types and group/cluster masses, as shown in Figs. 6$-$8,
which is described later.
It should be noted that moderate SF enhancement by ram pressure can  be
clearly seen only in the MW type (not so clear in the M33-type and the Dwarf-type 
for all models).

\subsubsection{Orbit}

The orbital pericenter distances ($r_{\rm p}$) are 42, 72, 102, 102, and 195 kpc
for the orbital types I1, I2, I3, B1, and B2 in the Virgo model. 
Accordingly, it is clear from Figs. 6 that 
$s_{\rm sf}$ can be  more
significantly reduced in the M33-type
with  smaller $r_{\rm p}$ among the three infall models (I1, I2, and I3).
This dependence on $r_{\rm p}$ can be seen in the M33-type and the Dwarf-type
for the two bound orbits.  Although final $s_{\rm sf}$ for the MW-type
is significantly smaller
than that of the isolated model for all orbits,  the dependence of final $s_{\rm sf}$
on $r_{\rm p}$ is not so clear for the MW-type.
Smaller final $\epsilon_{\rm sf}$ 
and more significant SF reduction 
are likely to occur in the Dwarf-type with smaller $r_{\rm p}$.
These derived dependences  of SFHs, $s_{\rm sf}$, and $\epsilon_{\rm sf}$ can be
seen in the Coma model and the Group one.
These dependences on the orbits of galaxies are due largely to
the differences of the influences of ram pressure force on gaseous evolution
(thus on SFHs) between different models.

The differences in $t_{\rm p}$ (the epoch of pericenter passage) between
models with different orbits can be as large as $\sim 1$ Gyr. This means
that the total gas masses ($M_{\rm g}$) of disks  at $t_{\rm p}$
can be significantly different between different models thus that the restoring force
of gas disks can be different between the models at $t_{\rm p}$. 
Therefore the differences of $M_{\rm g}$ at $t_{\rm p}$ might be  
responsible for the dependences of the present results  on
the orbits of galaxies. However, the differences in $M_{\rm g}$ at $t_{\rm p}$
between models with different orbits are rather small. For example,
$M_{\rm g}$ in the isolated MW model is
$1.11 \times 10^{10} {\rm M}_{\odot}$
at $t_{\rm p} \sim 1.3$ Gyr  (corresponding to the VI-B1-MW-IN model)
and $1.08 \times 10^{10} {\rm M}_{\odot}$ 
at $t_{\rm p} \sim 2.4$ Gyr (VI-I1-MW-IN model).
These values can be roughly estimated from the time evolution of $M_{\rm ns}$
in Fig. 3 and the adopted larger initial $M_{\rm g}$ 
($1.2 \times 10^{10} {\rm M}_{\odot}$).
Therefore, the possible small difference ($\sim 3$\%) in the restoring force
of gas disks is highly unlikely to influence the time evolution of SF in disks under
ram pressure. 
We therefore consider that the derived dependences of the present results
on the orbits of galaxies are due largely to the differences of the influences of
ram pressure force  on gas disks between different models.

\begin{figure*}
\psfig{file=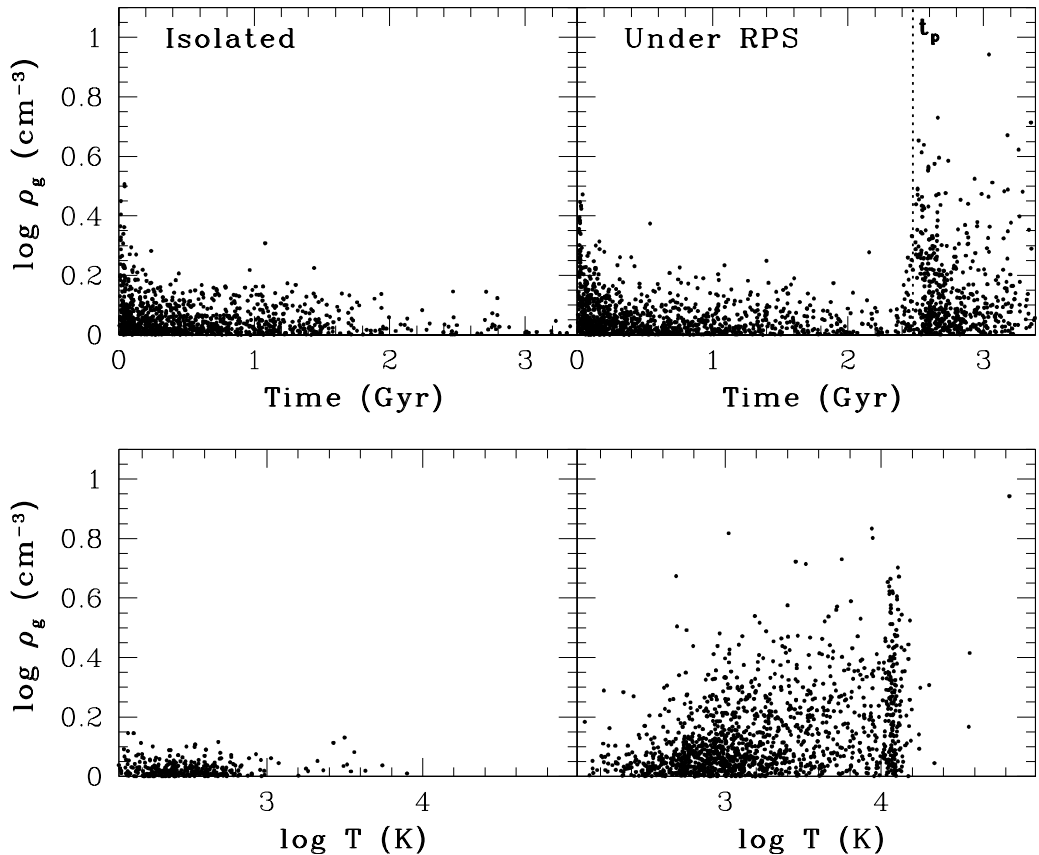,width=14.0cm}
\caption{
Upper panel: The time evolution of gas densities ($\rho_{\rm g}$ in units of H-atom number
per cm$^3$) for the isolated model and the model 
(VI-I1-MW-ED) under ram pressure stripping (right). 
The star formation histories of these models are already shown in Fig. 5.
These gas particles are those which are converted into new stars.
The file size of this figure can be very large ($\sim 7$ MB), if all gas particles
are plotted. Therefore, one every 20 particles is plotted in the upper panel
so that the file size can be much smaller. 
Nevertheless the differences in the the time evolution  of $\rho_{\rm g}$ 
between the two models
can be still clearly seen in the upper panel.
Lower panel: The locations of gas particles on the 
density-temperature ($\rho_{\rm g}-T$) plane
for the isolated model (left) and the VI-I1-MW-ED model (right).
The gas particles that are converted into new stars only after 
pericenter passage (2.48 Gyr) are shown (for the two models) 
so that the influences of ram pressure on the distributions of 
(star-forming) gas on the $\rho_{\rm g}-T$ plane can be more clearly
understood. 
One every 10 particles is plotted in this lower panel.
}
\label{Figure. 11}
\end{figure*}

\subsubsection{Group/cluster mass}

Fig. 7 shows that irrespective of galaxy-masses,  SF in disks can be suddenly and completely
truncated after pericenter passage in the Coma model, as long as $r_{\rm p}$ is small enough.
Not only the M33-type and the Dwarf-type but also  the MW-type shows   SF quenching  
in  the orbit I1  and I2 in this massive cluster model.
These results are in  a striking contrast with those for the low-mass Virgo model,
which implies that SF in MW-type disk galaxies can be quenched only in massive clusters.
It is intriguing that SF enhancement  does not occur for the MW-type
in I1 and I2 of the Coma model, 
though SF enhancement can be frequently seen in the Virgo model and 
the Group one. This result implies that moderately strong ram pressure is required
for SF of MW-type disk galaxies to be moderately enhanced. Too strong ram pressure
can simply strip the entire gas disk so that SF enhancement by compression of gas by
ram pressure can not occur. 

Fig. 8 shows that SF enhancement by ram pressure after pericenter 
passage in the MW-type is more pronounced
in  the Group model than in the Virgo one (in particular, orbits I1 and I2).
Furthermore, the final $\epsilon_{\rm sf}$ in the MW-type
is significantly larger than 1 for these
I1 and I2 orbits, which means that cold gas is more efficiently converted into new
stars owing to compression of gas by ram pressure.
These results imply that although $s_{\rm sf}$ of luminous disk galaxies
in a group can become slightly smaller
(i.e., slightly truncated H$\alpha$ disk)
owing to ram pressure of the intragroup gas,
their SF activity can be enhanced, as long as their $r_{\rm p}$ can be small enough.
The Dwarf-type with smaller $r_{\rm p}$ (I1 and I2 orbits) shows almost complete 
truncation of SF even in this group environment. 
It should be noted, however, that the DW-type with the B2 orbit
shows a very slight SF enhancement 
after pericenter passage (in comparison with the isolated model)  and thus
$\epsilon_{\rm sf}>1$. 
These results
suggest that ram pressure of intragroup gas can be important for the SFHs of disk
galaxies in group environments.

\begin{figure*}
\psfig{file=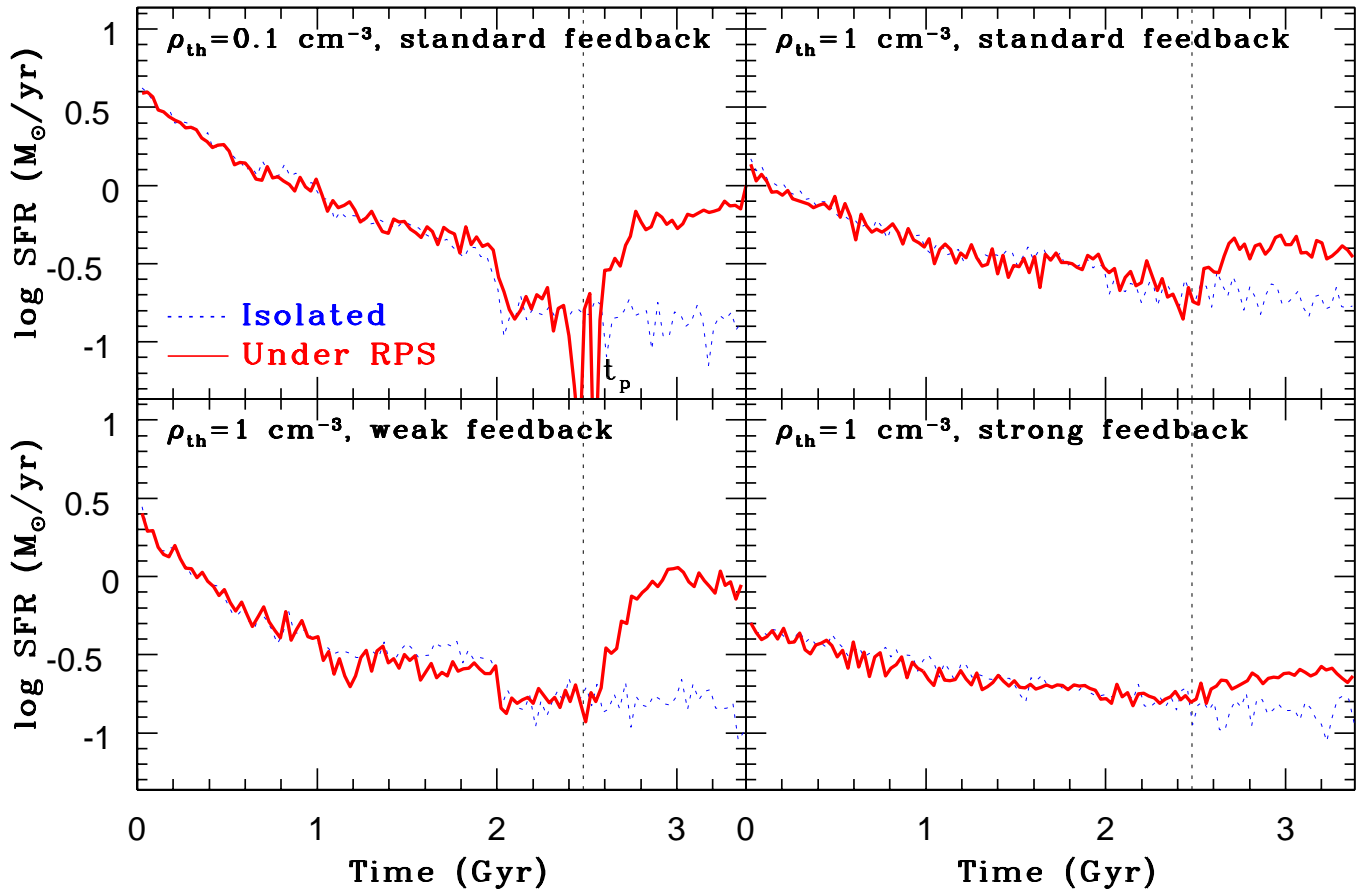,width=18.0cm}
\caption{
 The time evolution of SFRs 
for models with $\rho_{\rm th}=0.1$ cm$^{-3}$ and standard SN feedback  (upper left),
with $\rho_{\rm th}=1$ cm$^{-3}$ and standard SN feedback  (upper right),
with $\rho_{\rm th}=1$ cm$^{-3}$ and weak SN feedback effect  (lower left),
and with $\rho_{\rm th}=1$ cm$^{-3}$ and strong  SN feedback effect  (lower right).
Here weak, standard, 
and strong SN feedback models assume $t_{\rm adi}=3\times 10^5$ yr,
$10^6$ yr, and $10^7$ yr, respectively.
The black dotted line
indicates the time of pericenter passage ($t_{\rm p}$).
The isolated model  and the model under ram pressure stripping
are shown by blue dotted and red solid lines, respectively, in each frame.
}
\label{Figure. 12}
\end{figure*}

\subsection{Characteristic H$\alpha$ morphologies}

Fig. 9 presents a collection of the following four characteristic H$\alpha$
morphologies seen in the simulated disk galaxies of the present study:
(i) very compact and severely truncated, 
(ii) one-sided and outer crescent-like,
(iii)  broken ring-like and  patchy,
and (iv) an outer high-density ring (distributions). 
These characteristic H$\alpha$ morphologies are not seen in isolated models and thus
regarded as being due to ram pressure effects.
The severely truncated distribution can be seen 
more frequently in disk galaxy models  just before the complete
truncation of SF by ram pressure. The one-sided or crescent-like (or one-spiral-arm-like)
distribution can be more clearly  seen  in disk galaxy models with edge-on disk inclination
in which only one side of the gas disk can be strongly influenced by hydrodynamical
force of ram pressure.  Disk galaxies with these crescent-like H$\alpha$ distributions
often show enhanced global SFRs.

The broken ring-like distributions can be frequently seen in disk models with
face-on ($\theta=90^{\circ}$ and $\phi=0^{\circ}$) 
or inclined ($\theta=45^{\circ}$ and $\phi=30^{\circ}$)  disk configurations. 
This ring-like distribution is not due to the dynamical action of central bars,
because it is confirmed that this ring-like distribution can be seen in the disk models
with big bulges for which bar formation via global bar instability is unlikely.
The outer rings with significant SF enhancement within the rings can be seen
for  disk galaxy models  with smaller $r_{\rm p}$ orbiting  the Group.
The ring-like distributions of enhanced SF regions can not be seen in isolated models,
which suggests that such ring-like distributions can be used for searching for
evidence of an environmental effect of a group on SFHs of disk galaxies.

\subsection{Four modes of SF}

Fig. 10 summarizes the following four basic modes of
 SF in disk galaxies 
orbiting groups and clusters of galaxies: (i) Enhanced, (ii) Reduced,
(iii) Quenched, and (iv) Unaffected (SF).
The `Enhanced SF' mode means that SFRs can be moderately enhanced by ram pressure but
can not be as high as the SFRs of starburst galaxies  whereas
the `unaffected SF' mode means that SFRs can not change significantly 
(though very slight increase in SF can be seen during/after ram pressure).
The Enhanced SF mode is   more frequently seen in the MW-type
whereas the Quenched and Reduced SF modes are  more frequently seen in the
Dwarf-type in the present study.

As described in the preceding sections,
SFHs of disk galaxies can be enhanced or reduced depending on model parameters. 
Therefore, it would be unrealistic for any theoretical studies on statistical
properties of galactic SF to assume that
irrespective of galaxy-masses, group/cluster
masses, and orbital types,
SF of galaxies in groups and cluster
can be quenched after pericenter passage.
The origin of passive red spiral galaxies observed in distant clusters
(e.g., Dressler et al. 1998) can be closely related to the Quenched SF mode.
On the other hand, the origin of disk galaxies with local SF enhancement in
their outer disks observed in the Virgo cluster (e.g., Koopmann \& Kenney 2004)
has something to do with the Enhanced SF mode.

\subsection{A physical mechanism for enhanced SF}

Although we have shown that SF in disk galaxies can be enhanced by ram
pressure in the present simulations,
we have only briefly mentioned a possible physical process responsible
for this SF enhancement.
In order to demonstrate more clearly why ram pressure can cause
the modest enhancement of SF in disk galaxies,  we investigate
(i) the time evolution of gas densities (of SPH particles) and (ii)
the density-temperature diagram of the particles during and after
ram pressure stripping. We investigate these two points for gas particles
that are finally converted into new stars, because it is meaningless for us
to investigate these for all gas particles including stripped ones
for the purpose of clarifying the physical mechanism of the derived
SF enhancement. We choose the VI-I1-MW-ED mode for this investigation,
because the model clearly shows a SF enhancement. The SFH of this model
is already described  in Fig. 5 (as `Edge-on' model).

Fig. 11 shows that the densities of star-forming gas particles 
significantly increase after the pericenter passage of the disk galaxy
in this model. Fig. 11 furthermore shows that the locations of gas particles
on the $\rho_{\rm g}-T$ plane after the pericenter passage are quite different
from those in the isolated disk model. A significant fraction of gas particles
can have higher $\rho_{\rm g}$ and higher $T$ in the star-forming disk
under ram pressure. These results confirm that the modest SF 
enhancement  derived
in the present simulations is due to the significant increase of gas densities
caused by compression of gas by ram pressure of ICM.

\subsection{Dependences on SN feedback effects and SF threshold gas densities}

So far, we have presented the results of the models with a fixed $\rho_{\rm th}$
(=1  cm$^{-3}$) and $t_{\rm adi}$  (=10$^6$ yr). Although we consider
that these parameter values for SF threshold gas density
and SN feedback effect are quite reasonable and realistic for the present
investigation,  it is important for the present study to discuss whether or not
the present results are sensitive to the choice of these parameters.
We therefore discuss this point
by presenting the results for 
the models (VI-I1-MW-IN) with different $\rho_{\rm th}$ 
($=0.1$  cm$^{-3}$ and $1$  cm$^{-3}$) and $t_{\rm adi}$
($3\times 10^5$ yr, $10^6$ yr and $10^7$ yr,
which are  referred to as `weak', `standard', and `strong' SN feedback, respectively).
The isolated models with these weak and strong SN feedback 
show significantly higher and lower initial SFRs than a reasonable SFR of
$\sim 1 {\rm M}_{\odot}$ for a MW-type disk galaxy, respectively.
These models are therefore less realistic than those with the standard
SN feedback ($t_{\rm adi}=10^6$ yr).

Fig. 12 shows that modest enhancement in SF after pericenter passage
can be seen in the models with $\rho_{\rm th}=0.1$ and 1 cm$^{-3}$,
which confirms that the present results on the SF enhancement by ram pressure
are not sensitive to the choice of $\rho_{\rm th}$.
Furthermore, Fig. 12 shows that SF enhancement can be seen in the
model with strong SN feedback effects in which SN explosions can
cause gaseous outflow from the disk galaxy.  The SF enhancement
can be clearly seen in the models with different $t_{\rm adi}$, though
the levels of SF enhancement are slightly different between the three models
with weak, standard, and strong SN feedback.
These results demonstrate
that the present results on SF enhancement by ram pressure 
do not depend strongly on the details of the modeling
of SN feedback effects.

\section{Discussion}

\subsection{SF enhancement in  MW-type galaxies  under  ram pressure}

We have shown that {\it in some models},  SF can be 
enhanced rather than truncated in MW-type disk galaxy when they pass the pericenter 
of their orbits within clusters. The high pressure of ICM can (externally) compress
the cold ISM in disk galaxies so that SFRs, which are assumed to depend on local
gas densities in the present study, can significantly increase.  Using analytic models
of galaxy evolution in clusters,  Fujita \& Nagashima (1999) predicted that
SFRs of disk galaxies can increase by a factor of $\sim 2$ 
as galaxies approach the central regions of clusters,
because ISM of the galaxies can be strongly compressed 
by high pressure of ICM  at the pericenter
passages.  Previous numerical simulations also demonstrated that star formation
in MW-type  galaxies can be enhanced, if their ISM  are under strong influences
of high pressure of ICM (e.g., Bekki \& Couch 2003; Kronberger et al. 2008). 
It should be stressed, however, that 
this SF enhancement in MW-type disk galaxies is modest (i.e., not like a starburst
with a factor of $\sim 10$ SFR increase) and can not last long. 
Such a modest level of SF enhancement 
in disk galaxies under ram pressure
has been recently observed
by integrated-field spectroscopy and multi-band imaging
(e.g., Merluzzi et al. 2013).

The derived modest SF enhancement implies that 
SF in MW-type disk galaxies could not monotonically decline 
after they enter into cluster environments. Stripping of {\it halo gas}  of 
disk galaxies by cluster tide (Bekki et al. 2001)  and ram pressure  of ICM
(McCarthy et al. 2008; Bekki 2009) can cause gradual decline
of their star formation (e.g., Larson et al. 1980).
However,  moderately strong ram pressure on {\it ISM in disks} can enhance star formation
of disk galaxies in clusters: Cluster environments have positive and negative
effects on galactic star formation histories.
In recent semi-analytic  models (SAMs)
 of galaxy formation based on a $\Lambda$CDM cosmology,
star-formation quenching is assumed to occur in galaxies after their pericenter passages
within groups and clusters (e.g., Wetzel et al. 2013). 
The present study strongly suggest that (i) galactic  star formation in clusters
is not so simple as modeled in these semi-analytic studies,
and thus (ii) not only SF quenching but also
SF enhancement would need to be correctly included in their models for better
understanding galaxy evolution in clusters.

However, it is quantitatively unclear how dramatically the predictions
of SAMs can change if the above ram pressure effects on SFRs of galaxies
are more properly included in SAMs.  Given that the SF enhancement by
ram pressure is modest and a significant fraction of disk galaxies 
can have `unaffected' SF modes after ram pressure stripping,
such an inclusion of ram pressure effects in SAMs would not dramatically
change the already existing predictions. 
Since SF enhancement means more rapid gas consumption,
galaxies in groups/cluster might show reduced SF on average.

\subsection{Environmental dependences of H$\alpha$-to-optical-disk-size ratios}

One of key results in the present study is that 
the final H$\alpha$-to-optical-disk-size ratios ($s_{\rm sf}$) in disk galaxies 
under ram pressure of ICM depend primarily on
their orbits (in particular, pericenter distances),  
galaxy masses, inclination angles of disks
with respect to orbital directions,
and cluster masses.
Among these dependences of $s_{\rm sf}$,
an observable one is that   disk galaxies are likely to show  smaller $s_{\rm sf}$ 
in  more massive clusters. This dependence of $s_{\rm sf}$ on $M_{\rm h}$ would not be
so surprising and could be possibly confirmed by observations. 
However,  systematic observational studies on the dependence of $s_{\rm sf}$ on
$M_{\rm h}$ (or other cluster properties) have not been done yet. 
Koopmann \& Kenny (2004) investigated H$\alpha$ morphologies of spiral galaxies
in the Virgo cluster and found that about 52\% of the spiral galaxies show
truncated H$\alpha$ disks (smaller $s_{\rm sf}$), particularly, in those
with truncated H~{\sc i} disks (See their Fig. 15).
They therefore concluded that the observed  truncated SF regions in the Virgo
can be  caused by ISM-ICM hydrodynamical interaction. 

Future observational studies on the dependences of $s_{\rm sf}$ on $M_{\rm h}$
(also on group/cluster sizes and densities) could be doubtlessly worthwhile  as follows.
Moss \& Whittle (2000) 
investigated SF regions of disk galaxies in eight low-redshift Abell clusters
and revealed that strong evidence for tidally induced star formation can be
more frequently found in higher local galaxy surface density. Bretherton et al. (2013)
have also found that disk galaxies with enhanced SFRs in their cluster samples 
show evidence of disturbance. These two observational studies imply that
galaxy interaction can be also important in changing  SF regions of galaxies in clusters
and thus is a key parameter that controls $s_{\rm sf}$.
Therefore, a key question here is how the relative importance of ram pressure
and tidal interaction in determining $s_{\rm sf}$
changes with galaxy environments (e.g., $M_{\rm h}$).

The present study has shown that $s_{\rm sf}$  can be smaller for disk galaxies within
more massive clusters, because ram pressure is likely to be stronger in the clusters. 
Also the reduction of $s_{\rm sf}$ can be  less severe in groups  than in clusters
owing to the the weaker ram pressure force in groups.
On the other hand,  $s_{\rm sf}$ is unlikely to be less than 1 during tidal interaction, 
because both gas and old stars within a certain tidal radius can be equally stripped. 
Furthermore slow yet strong galaxy interaction can be an important physical mechanism 
for enhancing rather than quenching SF in
disk galaxies in groups  (Bekki \& Couch 2011).
Therefore, it would be possible that (i) more massive clusters show a larger number of
galaxies with smaller $s_{\rm sf}$ in comparison with less massive  clusters
and (ii) groups have different $s_{\rm sf}$ distributions in comparison with clusters
(e.g., most of group member galaxies show $s_{\rm sf}$ almost equal to or larger than 1).
Thus,  future observational studies on the dependences of $s_{\rm sf}$
on group/cluster masses ($M_{\rm h}$)
will  enable us to understand the relative importance of 
ram pressure and tidal interaction in determining
the 2D distributions of SF regions of disk galaxies in different environments.

\subsection{Formation of disky E+As}

The present study has shown that (i) low-mass disk galaxies can lose the entire gas disks
through ram pressure stripping at their pericenter passage even in low-mass clusters
with $M_{\rm h}=10^{14} {\rm M}_{\odot}$ and (ii) SF can be completely
and suddenly truncated (or quenched) after the pericenter passage. 
SF is very active in these low-mass, gas-rich disk galaxies before
SF quenching, and the stellar disk components can not be influenced at all by ram pressure
stripping of ISM in these galaxies. Therefore, the present results imply that
these low-mass disk galaxies which suddenly lose their gas can be identified as
`E+A' disk galaxies with poststarburst spectra $0.1-1$ Gyr after SF quenching.
Recent observational studies of spatial  distributions 
of stellar populations and kinematics in E+As 
(e.g., Pracy et al. 2009; 2012) have shown 
that the observed properties of most E+As are consistent with the simulated ones
based on 
a major merger scenario in which E+As can form from major merging with strong starbursts
and evolve then into elliptical galaxies (e.g., Bekki et al. 2005).
These results accordingly imply that disky E+As (i.e., those which are not formed from 
major merging) could be a minor population.

Although the observed negative radial gradients of Balmer absorption lines (i.e., stronger in
inner regions) are demonstrated to be consistent with the major merger scenario of
E+A formation (Pracy et al. 2012),
some E+As show rather flat radial gradients of Balmer absorption lines
(e.g., E+A 1 in Pracy et al. 2012).  Disk galaxies which very quickly lose their entire gas
disks through ram pressure stripping 
should show no/little radial gradients of Balmer absorption lines owing to sudden
and complete truncation of SF for the entire disks. Therefore,
the observed flat radial gradients of E+As in Pracy et al. (2012) appear to be consistent
with E+A formation through ram pressure stripping in disk galaxies.
Interestingly, E+A 1 in Pracy et al. (2012) has a dwarf-like
luminosity ($M_{\rm r} =-16.7$ mag), which appears to be consistent with the present result
that low-mass disks are more likely to become disky E+As.
Given that Tran et al. (2003) found many disky E+As in clusters,
the present study suggests that future observational studies on 2D distributions
of Balmer absorption lines in disky E+As will enable us to discuss whether
ram pressure stripping  can be a major mechanism for E+A formation in clusters.

\begin{figure*}
\psfig{file=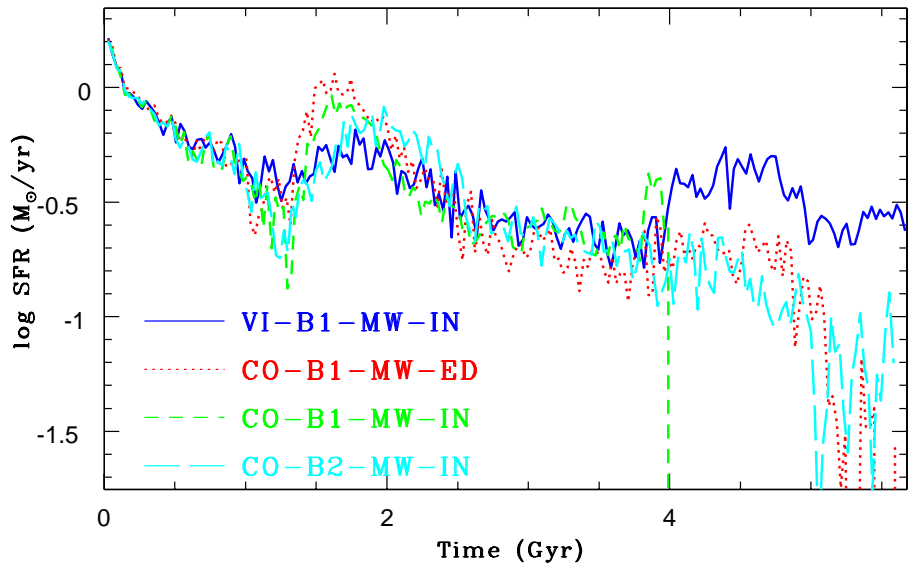,width=12.0cm}
\caption{
The long-time time evolution of  SFRs in the  
four models, 
V1-B1-MW-IN (blue, solid),
CO-B1-MW-ED (red, dotted), 
CO-B1-MW-IN (green short-dashed), 
and CO-B2-MW-IN (cyan, long-dashed).
For these models, the MW-type disk galaxies can pass through orbital pericenter
twice within the 5.6 Gyr orbital evolution.
The CO-B1-MW-IN model shows a significant enhancement before SF quenching around T=4 Gyr.
This SF mode of `Enhancement before quenching' can be seen in some models, though it
is not a major mode.
}
\label{Figure. 13}
\end{figure*}

\subsection{Timescale for SF quenching}

The present study has clearly shown that 
SF quenching (i.e., complete truncation of SF)
is possible in disk galaxies within groups and cluster for some models
under some physical conditions (e.g., smaller $r_{\rm p}$ and lower galaxy masses).
SF quenching has been demonstrated to occur in disk galaxies 
{\it at first pericenter passage} within clusters, as long as $r_{\rm p}$
is small enough ($r_{\rm p}$ well less than $r_{\rm s}$). 
Therefore, if the timescale for SF quenching is the time lag
between when a galaxy first enters within a virial radius of its host cluster
and when it passes through the orbital pericenter,
then the timescale can be $2-3$ Gyr for clusters.
The present study has also shown that SF in the MW-type disk galaxies
can not be completely truncated within 3.4 Gyr
(during just once pericenter passage),
if they have  bound orbits (B1 and B2) with
larger $r_{\rm p}$ in the high-mass clusters.
It is, however, unclear whether SF in MW-type disk galaxies within clusters
can be quenched if the galaxies can pass the
pericenter twice or more.

We have investigated the long-term SFHs (5.6 Gyr) 
of disks galaxies with bound orbits (B1 and B2) for
the Virgo model and the Coma one in order to understand whether SF quenching
is possible for the MW-type disk galaxies with bound orbits
and larger $r_{\rm p}$ in clusters. 
Fig. 13 shows that although SF quenching is not possible in the MW-type galaxies
with bound orbits within 5.6 Gyr 
for the low-mass Virgo cluster model, dramatic SF reduction (almost SF quenching)
is possible in the galaxies for the high-mass Coma model.
This result implies that (i) SF quenching by ram pressure stripping
is unlikely to  occur in luminous disk galaxies
for low-mass clusters and (ii) SF quenching timescale can be as long as $\sim 6$ Gyr
for luminous disk galaxies with larger  $r_{\rm p}$ for high-mass clusters.
Ram pressure stripping of cold gas from disk galaxies would not be the only
mechanism of SF quenching within several Gyr for clusters, and tidal stripping
due to galaxy interaction and cluster tide
could be also responsible for SF quenching.  It is thus our future study to
investigate how ram pressure and tidal interaction can combine to quench
SF in disk galaxies within clusters.

\section{Conclusions}

We have investigated the long-term SFHs (more than $\sim 3$ Gyr)
and the  spatial distributions of SF regions
of disk galaxies under strong ram pressure of ICM in clusters of galaxies by using
numerical simulations with a new model for time-varying ram pressure force.
The key parameters are (i) group/cluster halo masses ($M_{\rm h}$), 
(ii)  pericenter distances of orbits ($r_{\rm p}$),
(iii) inclination angles of disks with respect to the direction of their orbits
($\theta$ and $\phi$),
and (iv) galaxy-types (e.g., MW-type). The principal results of the present
numerical simulations are as follows. \\

(1) Ram pressure of ICM has the following three major effects 
on SFHs of disk galaxies in groups and clusters:
(i) temporary and moderate  enhancement of SF at pericenter passage,
(ii) significant decline of SFRs after pericenter passage,
(iii) complete truncation of SF (`SF quenching') after pericenter passage.
It depends on $M_{\rm h}$, $r_{\rm p}$, and $\theta$ ($\phi$) which of
these three effects can be the most dominant for SFHs of galaxies in groups and 
clusters. 
It would be therefore  unrealistic to adopt an assumption
that SF of  galaxies in groups and clusters
can be uniformly shut down after pericenter passage in theoretical
modeling on statistical properties of galactic SF and spectrophotometric properties
in groups and clusters. \\

(2) SF enhancement by ram pressure
is more likely to occur in MW-type disk galaxies,
and the degree of SF enhancement is more significant in the galaxies with lower
inclination angles (e.g., edge-on configurations). 
However, such SF enhancement is only temporary (during
and shortly after  pericenter passage) and
only modest (not like a factor of $\sim 10$ increase in SFRs) in most cases.
Less massive disk galaxies are less
likely to show SF enhancement in clusters, because the entire gas disk can be stripped
from less massive disks by even moderately strong ram pressure. 
SF enhancement by ram pressure can occur preferentially in the inner regions
of disks, which implies that the inner mass densities can be slightly
increased in disk galaxies under  ram pressure. \\

(3) The dominant role of ram pressure in galactic SFHs
is the significant reduction  of overall SF in disk
galaxies. The degree of SF decline depends strongly on group/cluster masses
for a given
set of galaxy parameters such that it can be more remarkable in groups and clusters
with larger $M_{\rm h}$. The level of SF decline by ram pressure can be more remarkable
in less massive disk galaxies for a given group/cluster mass.
Galaxies with smaller $r_{\rm p}$ show more severe SF decline 
in groups and clusters. \\

(4) Complete truncation of SF (`quenching') can occur in disk galaxies
with different masses after pericenter passage in high-mass clusters 
($M_{\rm h}=10^{15} {\rm M}_{\odot}$), as long as $r_{\rm p}$ is small enough
($<r_{\rm s}$). In low-mass clusters with $M_{\rm h}=10^{14} {\rm M}_{\odot}$,
SF quenching  is unlikely to occur in  MW-type disk galaxies owing
to weaker ram pressure.
These results imply that although the formation of anemic  spirals is possible
in low-mass clusters like the Virgo,  the formation of passive spirals
with no ongoing SF is unlikely in such clusters. \\

(5) Complete and sudden truncation of SF after pericenter passage is more likely to
occur in dwarf-like disk galaxies within groups and clusters. The likely outcome of this
is that the galaxies can show poststarburst (`E+A') spectra characterized
by strong Balmer absorption lines in the entire disks.
The possible no/weak  radial gradients of Balmer absorption lines in 
disky E+As with rotational kinematics
is in striking contrast to the  negative radial gradients  
observed in E+As formed from major galaxy merging.
Thus, the present study predicts that (i)
ram pressure can be responsible for the formation of disky E+As observed
in clusters and (ii) disky E+As are likely to be  dwarf-like  disk galaxies. \\

(6) Ram pressure stripping of the outer gas disks of galaxies
can significantly reduce the H$\alpha$-to-optical-disk-size ratios
($s_{\rm sf}=R_{\rm sf}/R_{\rm s}$)
of the galaxies after pericenter passage. The level of $s_{\rm sf}$ 
reduction is likely to be more remarkable for disk galaxies in more massive
clusters. The $s_{\rm sf}$ reduction
also depends on $M_{\rm h}$ and $r_{\rm p}$ of disk galaxies for a given group/cluster mass
and  disk inclination angles ($\theta$ and $\phi$)
such that it can be more remarkable for smaller $M_{\rm h}$
and smaller $r_{\rm p}$.  Such $s_{\rm sf}$ reduction can be seen for disk galaxies
even in massive groups 
with  $M_{\rm h}=10^{13} {\rm M}_{\odot}$
in which intragroup halo  gas is  warm rather than hot 
($T_{\rm ICM} \sim 6 \times 10^6$ K). This result implies that
SF regions can be also truncated (to a lesser degree) in group environment. 
Gas stripping by ram pressure can be efficient in groups
and thus responsible for the lack of high H~{\sc i}
mass galaxies observed in some groups (e.g., Kilborn et al. 2009). \\

(7) The global H$\alpha$ morphologies corresponding to 2D distributions of SF regions
in disk galaxies
can be significantly changed by ram pressure. Disk galaxies under strong ram
pressure can often show characteristic H$\alpha$ morphologies,
such as ring-like, one-sided, crescent-like, and highly asymmetric distributions.
These results imply that 2D H$\alpha$ distributions of disk galaxies
can be used for revealing a physical effect of ram pressure 
on disk galaxies in groups and clusters.  \\

\section{Acknowledgment}
I (Kenji Bekki; KB) am   grateful to the referee  for  constructive and
useful comments that improved this paper.
Numerical simulations  reported here were carried out
on the three GPU clusters,  Pleiades, Fornax,and gSTAR
kindly made available by International Centre
for radio astronomy research
(ICRAR) at  The University of Western Australia,
iVEC,  and the Center for Astrophysics and Supercomputing
in the Swinburne University, respectively.
This research was supported by resources awarded
under the Astronomy Australia Ltd's
ASTAC scheme on Swinburne with support from the Australian government.
gSTAR
is funded by Swinburne and the Australian Government's
Education Investment Fund.
%KB is grateful to Cameron Yozin-Smith for his reading this paper and giving
%useful comments to me.
KB acknowledges the financial support of the Australian Research Council
throughout the course of this work.

\appendix

\section{Resolution test}

\begin{figure*}
\psfig{file=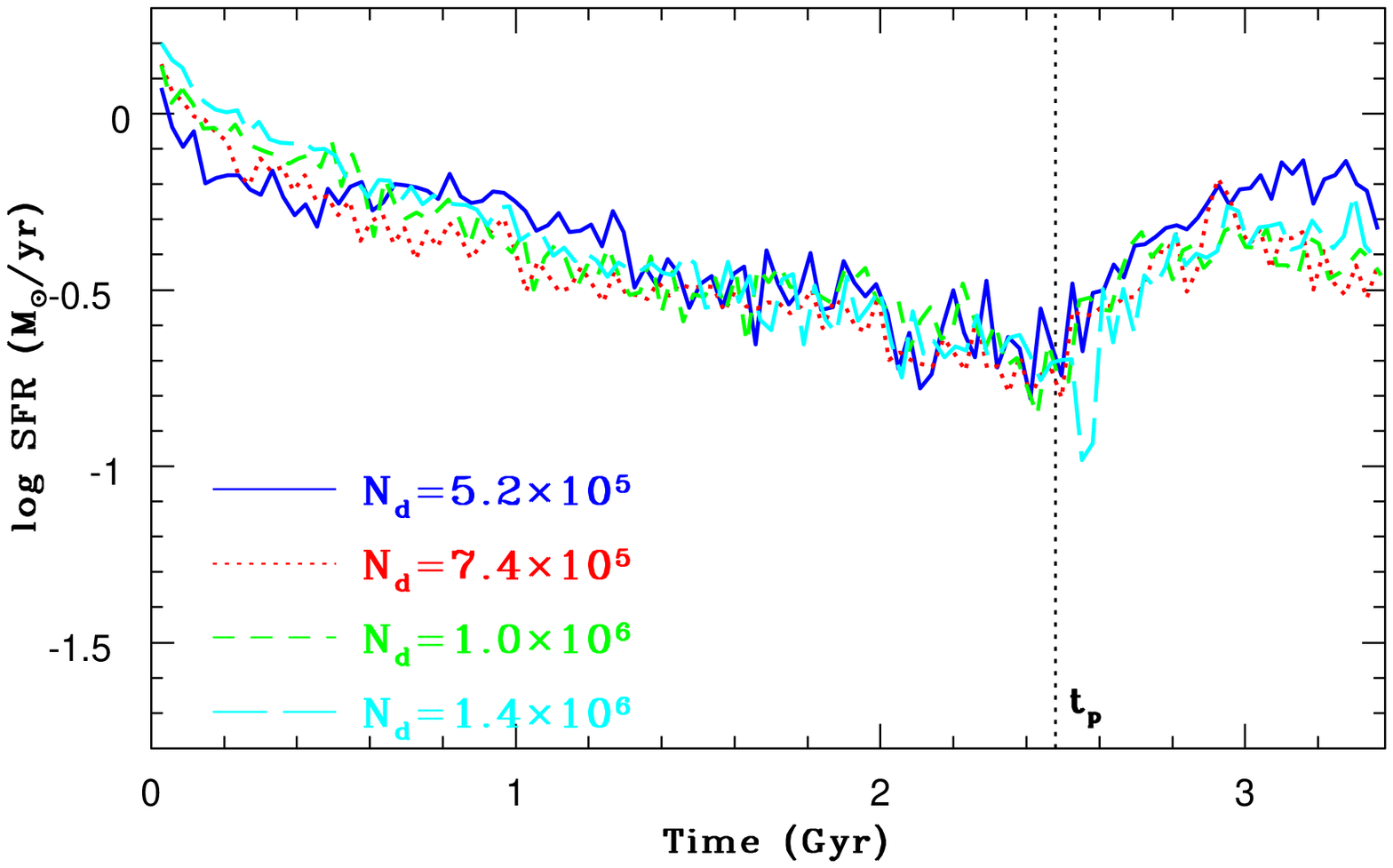,width=12.0cm}
\caption{
The time evolution of  SFRs in the MW-type disk galaxies with different
total number of particles used for a disk ($N_{\rm d}$) and the same other
parameters (VI-I1-MW-IN models):
$N_{\rm d}=2.1 \times 10^5$ (blue, solid),
$N_{\rm d}=5.2 \times 10^5$ (red, dotted),
$N_{\rm d}=1.0 \times 10^6$ (green short-dashed), 
and $N_{\rm d}=1.4 \times 10^6$  (cyan,
long-dashed). 
The time at the pericenter passage  is indicated by a black dotted line
and marked with `$t_{\rm p}$'.
}
\label{Figure. 14}
\end{figure*}

\begin{figure*}
\psfig{file=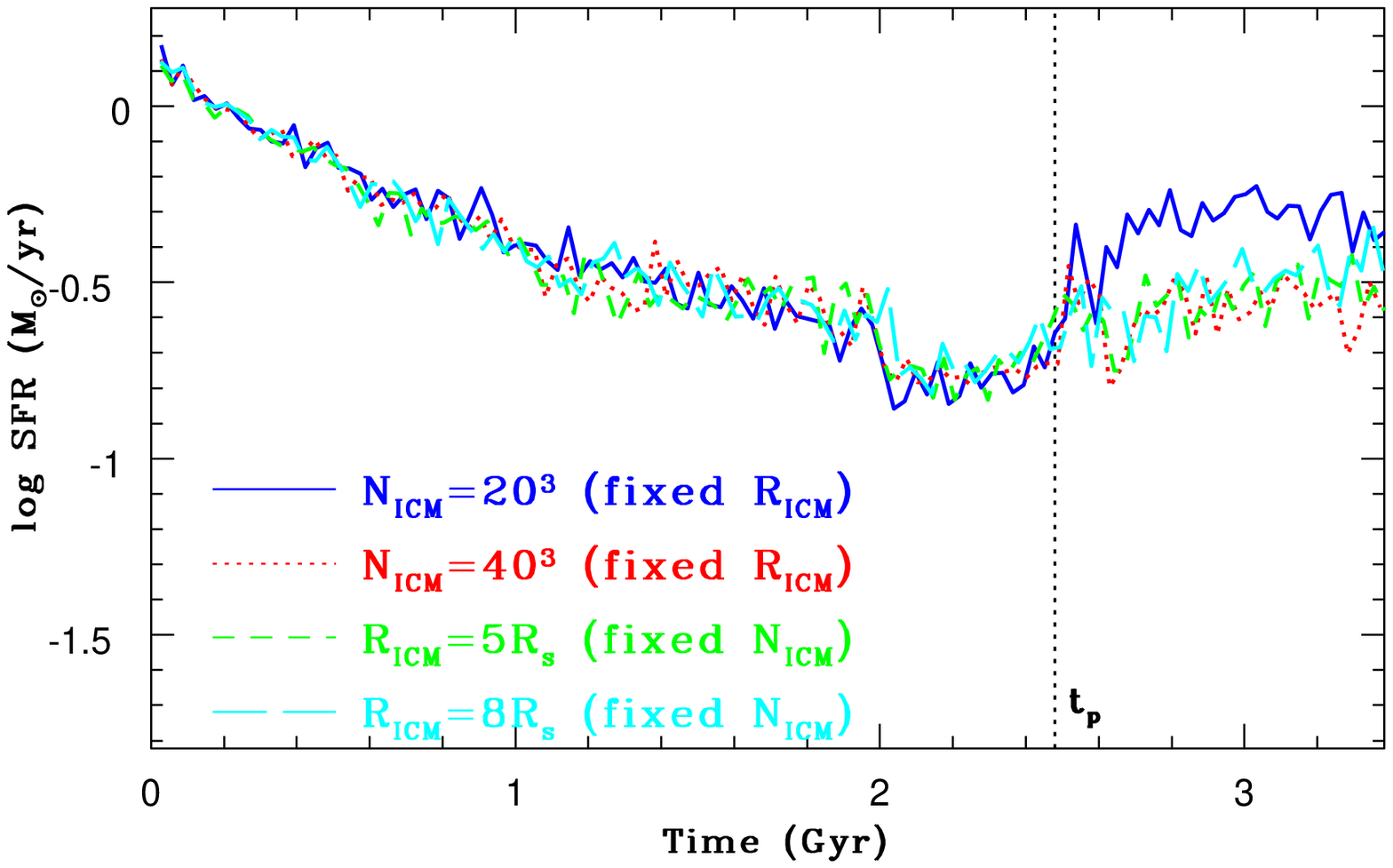,width=12.0cm}
\caption{
The time evolution of  SFRs in the MW-type disk galaxies with different
total number of ICM particles ($N_{\rm ICM}$) and the ICM box size
($R_{\rm ICM}$ in units of $R_{\rm s}$)  and the same other
parameters (VI-I1-MW-FA models):
$N_{\rm ICM}=20^3$ and $R_{\rm ICM}=6$ (blue, solid),
$N_{\rm ICM}=40^3$ and $R_{\rm ICM}=6$ (red, dotted),
$N_{\rm ICM}=40^3$ and $R_{\rm ICM}=5$ (green short-dashed), 
and $N_{\rm d}=40^3$  and and $R_{\rm ICM}=8$ (cyan,
long-dashed). 
The time at the pericenter passage  is indicated by a black dotted line
and marked with `$t_{\rm p}$'.
}
\label{Figure. 15}
\end{figure*}

The present numerical results on the effects of ram pressure of ICM on the SFHs of
disk galaxies could depend on the number of particles used for representing
disk galaxies and ICM. Also, the sharp boundary of a ICM box
could influence the details of hydrodynamical interaction between ICM and 
cold ISM in disk galaxies. Therefore, we here investigate how the present results
depend on (i) the number of particles in a disk galaxy ($N_{\rm d}$),
the number of ICM particles in a ICM box ($N_{\rm ICM}$), and the ICM box
size ($R_{\rm ICM}$) by using models with different $N_{\rm d}$,
$N_{\rm ICM}$, and $R_{\rm ICM}$.
The models in which disks are under the strong
influences of ram pressure are simply referred to as ram pressure models below.

The VI-I1-MW-IN model (`inclined' MW-type disk)
is used for investigating the dependences of the present
results on SF enhancement by ram pressure
in disk galaxies on $N_{\rm d}$,
because this model clearly shows SF enhancement after
pericenter passage.
The VI-I1-MW-FA model (`face-on' MW-type disk)
is used  for investigating how $N_{\rm ICM}$ and
$R_{\rm ICM}$ change the hydrodynamical interaction processes between ICM and ISM
and thus the SFRs, 
because the entire gas disk can be almost equally
influenced by ram pressure  force for this face-on model.
The results for the models with different $N_{\rm d}$
and with different $N_{\rm ICM}$ and $R_{\rm ICM}$ are shown in Figs. 14
and 15, respectively.

Fig. A1 clearly shows that 
SF can be moderately enhanced by ram pressure in the ram pressure models
for different $N_{\rm d}$.
Furthermore, the time evolution of SFRs is quantitatively similar between
the four different ram pressure models,
though the model with  $N_{\rm d} = 5.2 \times 10^5$
shows a more strongly enhanced SF after pericenter passage.
These results mean that as long as $N_{\rm d}$ is larger than $\sim 7 \times 10^5$, 
the present results on SFHs of disk galaxies in ram pressure models
do not depend on $N_{\rm d}$.

Fig. A2 shows that 
the time evolution of SFRs after pericenter passage is very similar between
the models, except the low-resolution one with $N_{\rm ICM}=20^3$. More massive
individual ICM particles in this low-resolution model ($N_{\rm ICM}=20^3$)
appear to compress more strongly the cold ISM
of the disk and consequently  enhance more significantly the SF.
These results mean that the dependences of the present results
on $N_{\rm ICM}$ and $R_{\rm ICM}$ are rather weak, as long as
$N_{\rm ICM} \ge  40^3$ and $R_{\rm ICM} \ge 5R_{\rm s}$.
Thus we consider that the present results for the models with
$N_{\rm d} \sim 10^6$, $N_{\rm ICM}=60^3$, and $R_{\rm ICM}=6R_{\rm s}$
can be regarded as robust.


\begin{thebibliography}{}

\bibitem[]{}
Abadi, M. G., Moore, B., Bower, R. G.,
1999, MNRAS, 308, 947

\bibitem[]{}
Balogh, M. L., Navarro, J. F., Morris, S. L., 2000, ApJ, 540, 113

\bibitem[]{}
Bekki, K. 2009,  MNRAS, 399, 2221

\bibitem[]{}
Bekki, K., 2013, MNRAS, 432, 2298

\bibitem[]{}
Bekki, K., Shioya, Y., 1998, ApJ, 497, 108

\bibitem[]{}
Bekki, K., Couch, W. J., Shioya, Y., 2001, PASJ, 53, 395

\bibitem[]{}
Bekki, K., Couch, W. J., Shioya, Y.,
2002, ApJ, 577, 651

\bibitem[]{}
Bekki, K., Couch, W. J., 2003, ApJ, 596, L13

\bibitem[]{}
Bekki, K., Couch, W. J., Shioya, Y., Vazdekis, A., 2005, MNRAS, 359, 949

\bibitem[]{}
Bekki, K.,  Owers, M. S.,  Couch, W. J., 2010, ApJ, 718, L27

\bibitem[]{}
Bekki, K.,  Couch, W. J., 2011, MNRAS, 415, 1783


\bibitem[]{}
Boselli, A.,  Gavazzi, G., 2006, PASP, 118, 517

\bibitem[]{}
Bretherton, C. F.,  Moss, C.,  James, P. A., 2013, A\&A, 553, 67

\bibitem[]{}
Brough, S., et al., 2013, MNRAS in press (arXiv:1308.2985)

\bibitem[]{}
Byrd, G. Valtonen, M., 
1990, ApJ,  350, 89

\bibitem[]{}
Cortese, L., et al., 2012, A\&A, 544, 101

\bibitem[]{}
Croom, S. M., et al., 2012, MNRAS, 421, 872

\bibitem[]{}
Crowl, H. H., Kenney, J. D. P., 2008, AJ, 136, 1623

\bibitem[]{}
Dressler, A., Oemler, A.,  Couch, W. J., Smail, I., Ellis, R. S.,
Barger, A., Butcher, H., Poggianti. B. M.,   Sharples, R. M.,
1997, ApJ,  490, 577

\bibitem[]{}
Evrard, A. E., 1991, MNRAS, 248, p8

\bibitem[]{}
Fossati, M., et al. 2013, A\&A, 553, 91

\bibitem[]{}
Fujita, Y.,  Nagashima, M., 1999, ApJ, 516, 619

\bibitem[]{}
Gunn, J. E., Gott, J. R. III., 1972, ApJ, 176, 1

\bibitem[]{}
Helsdon, S. F., Ponman, T. J., 2003, MNRAS, 339, L29


\bibitem[]{}
J\'achym, P., Palous, J., K\"oppen, J., Combes, F., 2007, A\&A. 472, 5

\bibitem[]{}
Kennicutt, R. C. 1998, ApJ, 498, 541

\bibitem[]{}
Kilborn, V. A., Forbes, D. A., Barnes, D. G., Koribalski, B. S.,  Brough, S., Kern, K.,
2009, MNRAS, 400, 1962

\bibitem[]{}
Koopmann, R. A., Kenney, J. D. P., 2004, ApJ, 613, 866

\bibitem[]{}
Kronberger, T., Kapferer, W., Ferrari, C., Unterguggenberger, S., Schindler, S.,
2008, A\&A, 481, 337

\bibitem[]{}
Larson, R. B., Tinsley, B. M.,  Caldwell, C. N., 1980,
ApJ, 237, 692

\bibitem[]{}
McCarthy, I. G., Frenk, C. S., Font, A. S., Lacey, C. G., Bower, R. G.,
Mitchell, N. L., Balogh, M. L., Theuns, T., 2008, MNRAS, 383, 593


\bibitem[]{}
Mamon, G. A., 1992, ApJ, 401, L3

\bibitem[]{}
Mastropietro, C.,  Burkert, A.,  Moore, B., 2009, MNRAS, 399, 2004

\bibitem[]{}
Matsumoto, H., Tsuru, T. G., Fukazawa, Y.,  Hattori, M., Davis, D. S.,
2000, PASJ, 52, 153

\bibitem[]{}
Merluzzi, P., et al. 2013, MNRAS, 429, 1747

\bibitem[]{}
Moore, B., Katz, N., Lake, G., Dressler, A., Oemler, A., 1996, Nat, 379, 613

\bibitem[]{}
Mori, M.,  Burkert, A., 2000, ApJ, 538, 559

\bibitem[]{}
Moss, C., Whittle, M., 2000, MNRAS, 317, 667

\bibitem[]{}
Navarro, J. F., Frenk, C. S.,  White, S. D. M.,
1996, ApJ, 462, 563 (NFW)

\bibitem[]{}
Neto, A. F., 2007, MNRAS, 381, 1450


\bibitem[]{}
Ponman, T. J., Bourner, P. D. J., Ebeling, H., B\"ohringer, H., 1996, MNRAS, 283, 690

\bibitem[]{}
Pracy, M. B., Kuntschner, H., Couch, W. J., Blake, C., Bekki, K., Briggs, F.,
2009, MNRAS, 396, 1349

\bibitem[]{}
Pracy, M. B., Owers, M. S., Couch, W. J., Kuntschner, H., Bekki, K., Briggs, F.,
Lah, P.,  Zwaan, M., 2012, MNRAS, 420, 2232

\bibitem[]{}
Quilis, V., Moore, B.,  Bower, R., 2000, Sci, 288, 1617

\bibitem[]{}
Rasmussen, J.,  Ponman, T. J., 2007, MNRAS, 380, 1554

\bibitem[]{}
Roediger, E., Hensler, G., 2005, A\&A, 433, 875

\bibitem[]{}
Roediger, E., Br\"uggen, M., 2007, MNRAS, 380, 1399

\bibitem[]{}
Rosen, A.,  Bregman, J. N., 1995, ApJ, 440, 634

\bibitem[]{}
Schulz, S., Struck, C., 2001, MNRAS, 328, 185

\bibitem[]{}
Shioya, Y., Bekki, K., Couch, W. J., De Propris, R., 2002, ApJ, 565, 223


\bibitem[]{}
Sutherland, R. S., Dopita, M. A., 1993, ApJS, 88, 253

\bibitem[]{}
Thornton, K., Gaudlitz, M., Janka, H.-Th.,  Steinmetz, M.,
1998, ApJ, 500, 95


\bibitem[]{}
Tonnesen, S., Bryan, G. L., 2012, MNRAS, 422, 1609

\bibitem[]{}
Vollmer, B., Cayatte, V., Balkowski, C., Duschl, W. J., 2001, ApJ, 561, 708

\bibitem[]{}
Wetzel, A. R., Tinker, J. L., Conroy, C.,  van den Bosch, F. C., 2013, MNRAS, 432, 336


\end{thebibliography}
\end{document}